\documentclass[journal,twocolumn,twopage,10pt]{IEEEtran}

\usepackage{amsmath,amssymb,amsthm,amsfonts,amsopn,cite,mathrsfs,todonotes}
\usepackage[T1]{fontenc}
\usepackage{ifpdf}
\usepackage[english]{babel}
\usepackage[ruled]{algorithm2e}
\usepackage{dsfont,url}
\usepackage{tikz}
\usepackage{pgfplots}
\pgfplotsset{compat=1.14}
\usepackage{xcolor}
\usepackage{booktabs}

\newcommand*\justify{%
  \fontdimen2\font=0.4em
  \fontdimen3\font=0.2em
  \fontdimen4\font=0.1em
  \fontdimen7\font=0.1em
  \hyphenchar\font=`\-
}

\usepackage{multirow}
\usetikzlibrary{patterns}
\usetikzlibrary{calc}
\usetikzlibrary{arrows.meta}

\makeatletter

\usepackage{bm}
\usepackage{hyperref}

\newcommand{\RR}{\mathbb{R}}
\newcommand{\ZZ}{\mathbb{Z}}
\newcommand{\NN}{\mathbb{N}}
\newcommand{\CC}{\mathbb{C}}

\newcommand{\abs}[1]{|#1|}

\newcommand{\revised}[1]{#1}
\newcommand{\revisedtwo}[1]{#1}
\newcommand{\revthree}[1]{#1}

\usepackage{matlab-prettifier}
\usepackage[T1]{fontenc}
\lstMakeShortInline"
\begin{document}

%

\author{Nicki Holighaus,
				G\"unther Koliander, 
        Clara Hollomey, and Friedrich Pillichshammer
\thanks{Manuscript received XXX; revised August XXX.}%
\thanks{N.\ Holighaus (corresponding author) and C. Hollomey are with the
Acoustics Research Institute (ARI), Austrian Academy of Sciences,
Wohllebengasse 12--14, 1040 Vienna, Austria. G. Koliander is with ARI and 
the Faculty of Mathematics, University of Vienna, Austria. F. Pillichshammer
is with the Institute of Financial Mathematics and Applied Number Theory, Johannes Kepler University Linz, Austria. 
e-mail: 
 \texttt{\{nicki.holighaus,guenther.koliander,clara.hollomey\}}\linebreak{}
 \texttt{@oeaw.ac.at}, 
 \texttt{friedrich.pillichshammer@jku.ac.at}
}%
\thanks{Extended results, audio files and code for reproducing the presented experiments is available at: 
 \texttt{ltfat.org/notes/057}}%
\thanks{This work is supported by the Austrian Science Fund (FWF): I 3067–N30 (N.H.), Y 1199 ``Time-Frequency Analysis, Randomness and Sampling.'' (G.K.), and F5509-N26, which is a part of the Special Research Program ``Quasi-Monte Carlo Methods: Theory and Applications.'' (F.P.). 
}%
\thanks{Copyright (c) 2023 IEEE. Personal use of this material is permitted.
However, permission to use this material for any other purposes must be
obtained from the IEEE by sending a request to pubs-permissions@ieee.org.
}%
}

\markboth{Holighaus, Koliander, Hollomey, and Pillichshammer}{Wavelet lattices for audio}

 \title{\revthree{Grid-Based Decimation for Wavelet Transforms with Stably Invertible Implementation}}


\maketitle
\begin{abstract}%
The constant center frequency to bandwidth ratio (Q-factor) of wavelet transforms provides a very natural representation for audio data. However, invertible wavelet transforms have either required non-uniform decimation---leading to irregular data structures that are cumbersome to work with---or require excessively high oversampling with unacceptable computational overhead. 
\revthree{Here, we present a novel decimation strategy for wavelet transforms that leads to stable
representations 
with  oversampling rates close to one and uniform decimation. 
Specifically, we show that 
finite implementations of the resulting representation are  energy-preserving in the sense of frame theory.} The obtained wavelet coefficients can be stored in a time-frequency matrix with a natural interpretation of columns as time frames and rows as frequency channels. 
This matrix structure immediately grants access to a large number of algorithms that are successfully used in time-frequency audio processing, but could not previously be used jointly \revthree{with wavelet transforms.} 
We demonstrate the application of our method in processing based on nonnegative matrix factorization, in onset detection, and in phaseless reconstruction.\end{abstract}

\IEEEpeerreviewmaketitle
\section{Introduction}
The wavelet transform is one of the most important and well-studied time-frequency filter banks, only rivaled by the short-time Fourier transform. Due to its constant center frequency to bandwidth ratio, or \emph{Q-factor}~\cite{brown1992efficient}, wavelets provide a natural and physically meaningful representation of audio\revised{: The impulse response of every filter captures an equal number of wavelengths of its center frequency. 
The constant Q-factor ensures that equal intervals on (Western) musical scales, which consist of geometrically spaced tones, are resolved equally well, independent of their absolute position. Finally, at frequencies above $500$\,Hz, the peripheral auditory system in humans is well-approximated by a constant-Q filter bank~\cite{moore2012introduction}. 
For a short  discussion, see, e.g.,~\cite{schorkhuber2010constant}.} 
Consequently, constant-Q filter banks in general~\cite{schoerkhuber2013audio,fuentesCQT2012,TodiscoCQT2017} and wavelet transforms in particular~\cite{SAKAR2019255,119752,1021072,1495445} have been used to great success in the analysis of speech and audio signals. 
For the longest time, however, constant-Q filter banks have been inaccessible, or at least inconvenient, for processing audio at sufficient fidelity: 
Wavelet bases~\cite{Mallat2009Wavelet} and undecimated, discrete wavelet systems ~\cite{10.1007/978-3-642-75988-8_28,10.1007/978-3-642-75988-8_29} possess a very low Q-factor that is not suitable for processing audio. 
\revised{Modern implementations of constant-Q filter banks allow for the tuning of the Q-factor and the oversampling rate, but computationally efficient, numerically stable, and invertible constant-Q filter banks have only recently been constructed: In~\cite{SelesnickWavelet2011} this is achieved by purposeful modification of Mallat's algorithm for the fast wavelet transform, and in~\cite{dogrhove13,schoerkhuber2014a} by means of mathematical frame theory~\cite{Christensen2016}. 
To achieve perfect reconstruction, these constructions rely on non-uniform decimation, choosing sufficiently small decimation factors inversely proportional to the bandwidth or, equivalently, center frequency of the constant-Q filters.%
\footnote{\revised{This approach for choosing decimation factors is not exclusive to \emph{invertible} constant-Q transforms, but shared across most constant-Q and wavelet transform implementations.}} 
Consequently, every channel produces coefficients at different, usually unmatched, rates. 
Alternatively, it is possible to employ no in-channel decimation whatsoever, resulting, however, in prohibitively high oversampling at large Q-factors.

Although the availability of invertible representations with appropriate frequency resolution, i.e., a sufficiently large Q-factor, presents an important step towards making constant-Q filter banks more attractive for audio processing, another fundamental issue remains, previously discussed in~\cite{schorkhuber2010constant}, alongside a partial solution: Employing non-uniform decimation, the filter bank coefficients form an irregular data structure that is fundamentally more difficult to work with than matrices, both computationally and conceptually. Such coefficients are incompatible with any method that relies on matrix manipulation, and substantial work is required to integrate them into readily available audio processing toolboxes such as, e.g.,~\cite{pd} and~\cite{mirtoolbox}, usually implemented under the assumption of matrix-structured coefficients. 
This effect is most readily apparent in algorithms for the ubiquitous short-time Fourier transform (STFT), which often rely on the segmentation of the STFT coefficients into \emph{time frames}, see~\cite{mp3standard} and~\cite{gardner}.
These time frames are processed individually (or in small groups) and often in real-time, but the segmentation relies on the assumption that the coefficients are time-aligned. More precisely, the between channel coefficient relations are assumed to be position-independent, which is clearly not the case for non-uniform decimation. Hence, the adaptation of successful processing schemes for STFT coefficients to the constant-Q setting remains challenging. Nonuniform decimation also has implications concerning the reconstruction procedure itself, see Section \ref{ssec:wavelets}. Most importantly, even efficient reconstruction algorithms~\cite{necciari18,6637697} involve costly iteration, unless all filters are strictly band-limited~\cite{balazs2011theory}, precluding the use of finite impulse response filters.

In this work, we present a family of wavelet filter banks 
based on quasi-random sampling of the continuous wavelet transform that are computationally efficient, with tunable Q-factor through unrestricted choice of the mother wavelet, and flexible oversampling rate. With 
\revthree{ oversampling rates close to one,} the proposed filter banks achieve perfect reconstruction of discrete signals. Uniquely, they do so while employing uniform decimation and thereby overcoming the difficulties induced by the irregularity of previous invertible constant-Q transforms.
}


\vspace{6pt}
\noindent\textit{Previous work on random and quasi-random sampling of time-frequency representations:} This work is not the first to consider (quasi-)randomized sampling of time-frequency representations or more general integral transforms. In a series of recent works, Levie et al.\ consider Monte Carlo~\cite{levie2021randomized1,levie2021randomized} and Quasi-Monte Carlo~\cite{levie2021quasi} style random sampling of time-frequency integral operators. Their work is concerned with the approximation of continuous domain time-frequency processing by means of Monte Carlo integration with (quasi\nobreakdash-)random time-frequency samples. By invoking prior results on Monte Carlo integration, the authors demonstrate that the approximation error can be controlled when \revised{a technical structure condition} is satisfied. 
They further show that this condition is satisfied by the STFT, the wavelet transform, and a custom blending of the two, referred to as localizing time-frequency transform. The invertibility or stability of the sampled representation is not investigated, however. In the context of random sampling, these properties are considered in the literature on \emph{relevant sampling}, introduced by Bass and Gr\"ochenig for bandlimited functions~\cite{bass2013relevant} and later generalized to various settings~\cite{FUHR20191,PATEL2020124270,goyal2021random}, including time-frequency representations~\cite{velasco2017relevant}. Relevant sampling provides a probabilistic framework for stable sampling of functions that are localized in a domain of finite volume, e.g., bandlimited signals that have only negligible energy outside a finite interval. 
Our work differs from these prior works in multiple ways: Our proposed sampling sets are not fully (quasi-)random, but correspond to a uniform time-frequency grid, up to the introduction of a small, quasi-random delay in every wavelet channel. Further, we consider perfect reconstruction of arbitrary signals, without localization assumptions.    

\vspace{6pt}
\noindent\textit{Contribution: } 
\revthree{Previously, grid-based decimation strategies were considered ill-suited for the continuous wavelet transform and not expected to provide efficient and stably invertible  representations even for discrete signals. Indeed, known constructions did not admit perfect reconstruction at moderate, or even low, oversampling rates.} 
In this paper, we propose the first grid-based decimation strategy for wavelet transforms with a tunable Q-factor that allows perfect reconstruction at oversampling rates close to $1$ and that provides excellent numerical stability\revised{, in the sense of energy preservation,\footnote{\revised{For more details on stability and energy preservation, see Section \ref{ssec:wavelets}.}}} at moderate oversampling rates, ranging from $2$ to $8$, as commonly used for audio processing with the STFT. To achieve this, we combine shift-invariant systems~\cite{Janssen1998,BoelcskeiFilterbanks1998} with ideas from quasi-random sampling using low-discrepancy sequences~\cite{Niederreiter1992random,dick_pillichshammer_2010}. Similar to wavelet bases and prior invertible constant-Q implementations, we  \revised{use a set of compensation filters to cover} an arbitrarily small low-frequency region. Our construction is validated \revised{in the finite domain by computing exact frame bound ratios} and accumulated spectrograms of the decimated wavelet systems across a range of system parameters covering variations of the mother wavelet, the number of frequency channels, the decimation factor, and the oversampling rate. 
As a proof of concept, we apply the proposed wavelet decimation to several audio applications. 
We replicate an experiment on signal enhancement based on Itakura-Saito nonnegative matrix factorization (NMF) as proposed by F\'evotte et al.~\cite{fevotte2009nonnegative}, which relies heavily on the natural interpretation of the representation coefficients as a \emph{time-frequency matrix}. \revised{Furthermore, we illustrate the use of our proposed scheme in onset detection based on a straightforward adaptation of the classic spectral flux method~\cite[Sec.~3-A]{bello2005tutorial}}.
\revised{As an indication that the proposed scheme performs on par with established, non-uniform  constant-Q transforms in tasks that do not require matrix structure, we further evaluate the suitability of the proposed method for phaseless reconstruction with the fast Griffin-Lim algorithm~\cite{ltfatnote021}. In addition to the standard, irregular wavelet decimation, we also compare to the STFT.}

\vspace{6pt}
\noindent\textit{Paper structure: } 
A short introduction to wavelet systems and quasi-random sequences is given in Section~\ref{sec:background}, before presenting the proposed decimation scheme in Section \ref{sec:oursampling}. 
In Section~\ref{sec:computation}, we discuss details of practical implementation and complexity, and  evaluate our construction numerically.
We further apply the proposed scheme in three illustrating experiments in audio processing (Section~\ref{sec:experiments}). Specifically, we consider the decomposition of a signal with nonnegative matrix factorization, onset detection based on an adaptation of spectral flux, and phaseless reconstruction from time-frequency coefficients based on the fast Griffin-Lim algorithm. 
The paper concludes with a summary of the results and an outlook towards related future work (Section~\ref{sec:conclude}).

\section{Technical Background}\label{sec:background}

Before introducing our novel decimation strategy, we review some basics of wavelet systems and quasi-random sequences. 

\subsection{The Wavelet Transform}
\label{ssec:wavelets}

A wavelet system is a collection of functions (or vectors) generated from a single prototype, the \emph{mother wavelet}, by translation and dilation. Since we are interested in real-valued signals, audio signals in particular, we consider a mother wavelet $\psi$ such that its Fourier transform vanishes for negative frequencies, i.e., $\hat{\psi}(\xi) = 0$ for $\xi\in (-\infty,0]$. Such mother wavelets are often called \emph{analytic}, although the terminology \emph{analytic wavelet transform} has been used in at least two different manners in the past~\cite{ltfatnote053,Lilly2010analytic}. 
The continuous wavelet system is generated via dilation by $s>0$ and translation by $x \in \RR$ of the mother wavelet:
  \[
   \psi_{(x,s)} := s^{-1/2} \psi\bigg(\frac{\bullet-x}{s}\bigg).
  \]
A signal $f$ can now be filtered using this system resulting in the continuous wavelet transform
\begin{equation}\label{eq:waveletttransform}
    \begin{split}
    W_{\psi}f (x,s) 
    & = \frac{1}{\sqrt{s}} \int f(t) \overline{\psi\bigg(\frac{t-x}{s}\bigg)} \, \mathrm{d}t\\
    & = \langle f, \psi_{(x,s)} \rangle_{L_2} \\
    & = \big[f \ast \overline{\psi\left(-\bullet/s\right)}\big](x)\,.
    \end{split}
\end{equation}
The mathematical study of wavelet transforms usually considers $W_{\psi}f$ in terms of  the inner product representation of the transform coefficients, but the final equality justifies the interpretation as a filter bank. Importantly, the wavelet transform is shift-invariant, i.e., the wavelet transform of a delayed signal $f(\bullet-y)$ equals the delayed (in the first variable) wavelet transform of $f$. Also note that the scale $s$ is inversely proportional to the filter center frequency: If $\psi$ has its passband around frequency $\xi_1$, then $\psi(\bullet/s)$ has its passband around $\xi_s := \xi_1/s$. 

In applications, only a discrete subset of the continuous wavelet system can be considered; the system, or equivalently the transform, is \emph{decimated}. 
\revised{
Commonly, the discrete subset $\big((x_{l,j},s_j)\big)_{l\in\ZZ,j\in I}\subset \RR\times [0,\infty)$ of translation-dilation pairs is generated by certain decimation rules. 
We denote the corresponding decimated wavelet system by $\{\psi_{l,j}\}_{l\in\ZZ,j \in I}$ with $\psi_{l,j} := \psi_{(x_{l,j},s_j)}$.
Regarding the choice of decimation $\big((x_{l,j},s_j)\big)_{l\in\ZZ,j\in I}$, it is desirable that any function $f$ can be 
stably recovered from the decimated transform and the energy of $f$ is accurately represented by its coefficients.} 
Mathematically, these properties \revised{are equivalent} to the discrete subset \revised{$(\psi_{l,j})_{l,j}$} of the continuous wavelet system constituting a frame~\cite{Christensen2016}\revised{, i.e., it satisfies the energy equivalence relation
\begin{equation}\label{eq:frameineq}
  A\|f\|_2^2 \leq \sum_{l,j} |\langle f,\psi_{(l,j)}\rangle|^2 \leq B\|f\|_2^2,
\end{equation}
 for all $f$ and some constants $0<A\leq B<\infty$. Implicitly, we always assume that $A,B$ are the optimal constants such that \eqref{eq:frameineq} holds. The ratio $R_\textrm{FB} := B/A$ of the so-called upper ($B$) and lower ($A$)} frame bounds quantifies how well the decimated transform preserves signal energy, i.e., how stable it is in numerical computation. \revised{If $R_\textrm{FB}\neq 1$, then error-free reconstruction of $f$ from the decimated transform requires a \emph{dual frame}, which can be efficiently precomputed under certain conditions on mother wavelet and decimation~\cite{dagrme86,balazs2011theory,dogrhove13}, and realized by iterative schemes otherwise~\cite{gr93acceleration,necciari18,6637697}. If the transform coefficients are modified, the energy of the synthesized function $\tilde{f}$ can nevertheless be bounded by $A^{-1}$ times the energy of the modified coefficients. Note that the dual frame of a wavelet frame is not necessarily a wavelet frame, or even a filter bank. In fact, that is the exception rather than the norm.} 

\begin{figure}[!t]
\includegraphics[width=0.5\textwidth,trim=9cm 5.1cm 4cm 0cm, clip]{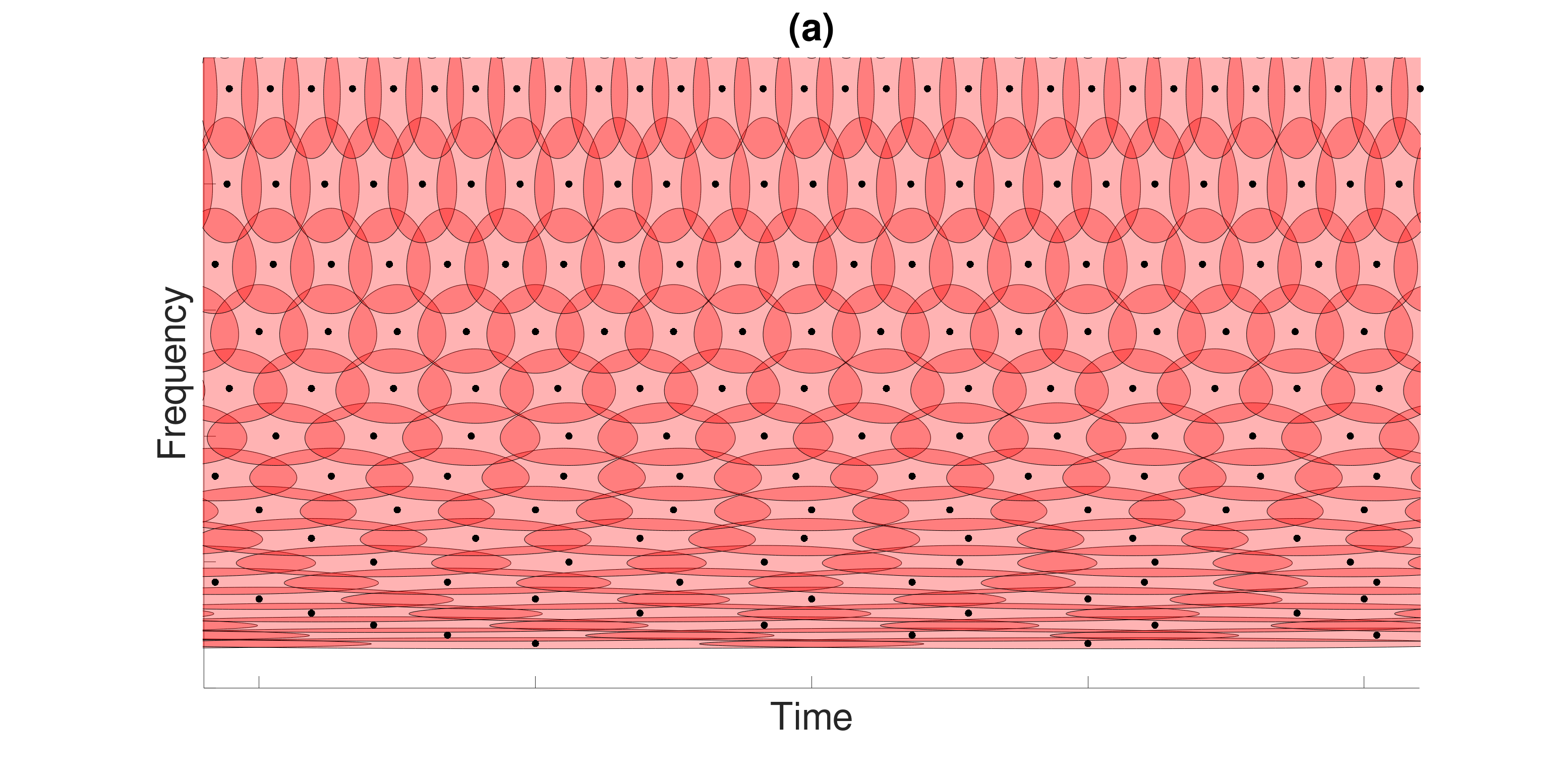}
\includegraphics[width=0.5\textwidth,trim=9cm 2.1cm 4cm 0cm, clip]{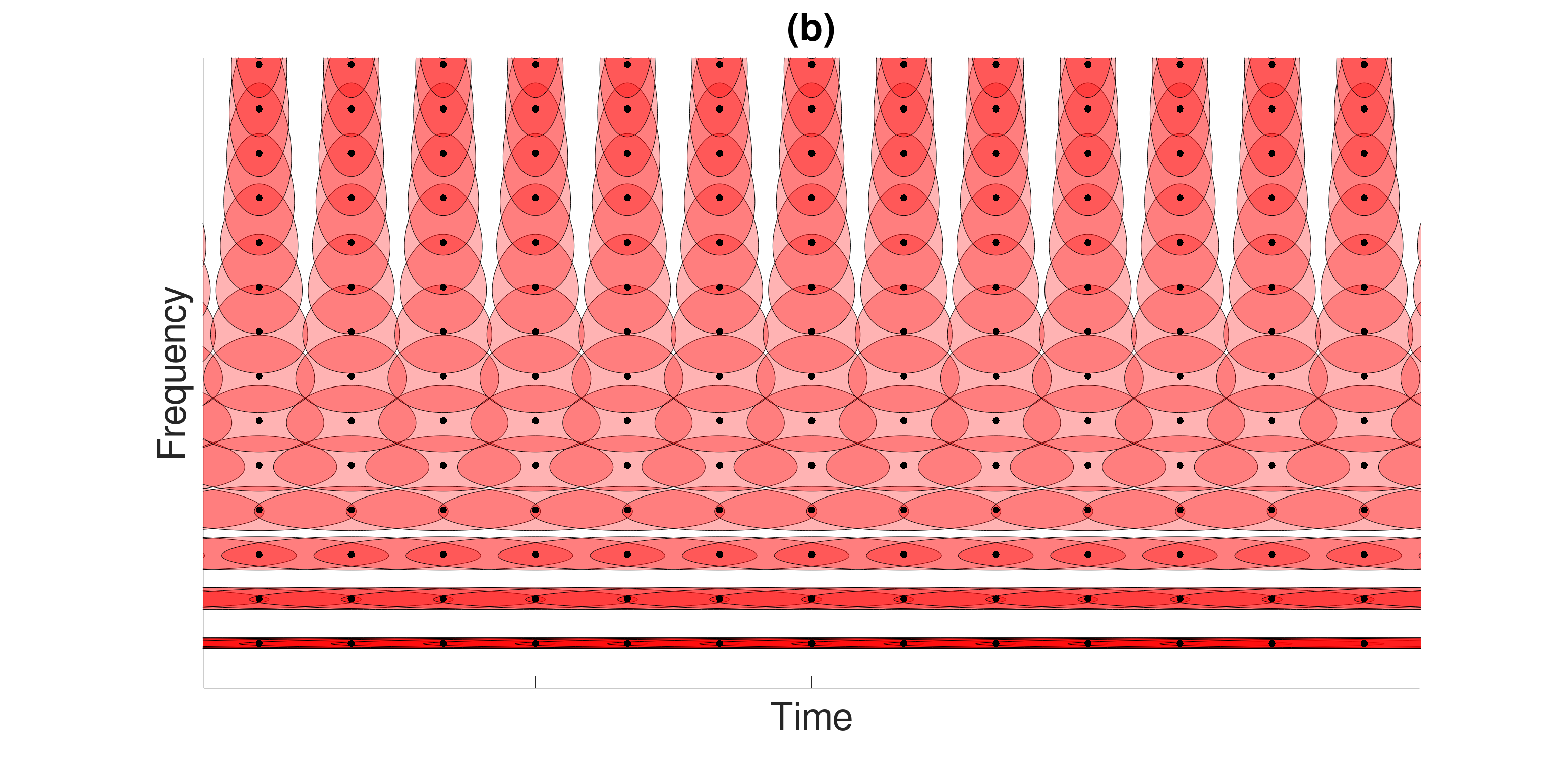}
\caption{Time-frequency geometry of decimated wavelet systems: (a) The classical nonuniform wavelet decimation provides a mostly even covering of time-frequency space, at the cost of introducing nonuniform decimation. (b) Wavelet decimation on a uniform grid results in an increasingly uneven covering away from a small frequency strip. \revisedtwo{The depicted hyperbolic circles have identical radius across both panels.}\label{fig:regularwavelets}}
\end{figure}

Regarding the choice of the scales $s_j$, recall that
the bandwidth of wavelet filters increases linearly with their center frequency $\xi_s$, which is inversely proportional to their scale $s$.
Thus, it is natural to decimate the frequency channels by taking integer powers of a fixed \emph{base scale} $a>1$, i.e., $s_j = a^{-j}$. 
Conversely, the width of the wavelet impulse responses is proportional to the scale, suggesting in-channel decimation according to $x_{l,j} = a^{-j}\cdot lb$ for some fixed $b>0$, see also \cite[Chapter 3]{da92}. 
Intuitively, the time-frequency region covered by the individual $\psi_{l,j}$ corresponds roughly to ellipses of constant area that grow narrower as $j$ increases, or, more accurately, to hyperbolic circles of constant radius,%
\footnote{\revised{A hyperbolic circle contains all time-scale pairs that have a hyperbolic distance less than a given radius to the center $(x_{l,j},s_j)$. 
Here, hyperbolic distance between $(x,s)$ and $(x',s')$ is measured as $2 \operatorname{arsinh} \frac{\sqrt{(x'-x)^2+(s'-s)^2}}{2\sqrt{ s s'}}$.
Note that we depict these circles in the time-frequency rather than the time-scale plane.}}
centered at $(x_{l,j},\xi_1/s_j)$, see Fig.~\ref{fig:regularwavelets}(a).

This decimation strategy choosing  $(x_{l,j},s_j) = (a^{-j}\cdot lb,a^{-j})$, with geometric frequency spacing and channel-dependent decimation factor (hop size) $d_j = a^{-j}\cdot b$, is customary and well-studied for wavelet bases, overcomplete wavelet systems, and 
 constant-Q transforms, e.g.,~\cite{da92,dogrhove13}. We will refer to this convention as \emph{classical wavelet decimation}. 
 \revised{If $\psi$ is not strictly bandlimited or the decimation factors are too large, then the dual frame required for error-free reconstruction will only be a filter bank in exceptional cases.}
Furthermore, the dependence of the decimation factor $d_j$ on the scale leads to filter bank coefficients that are not time-aligned, \revised{and, for a lack of position-independent relations between  coefficients in different bands, cannot be represented as a time-frequency matrix or segmented into time frames of identical structure. In many applications and implementations, this irregularity encumbers the workflow. }
In contrast to the STFT, the time-frequency geometry of the wavelet transform described above is ill-suited for decimation on a uniform, rectangular grid, independent of whether the center frequencies are spaced linearly or logarithmically, see Fig.~\ref{fig:regularwavelets}(b). It is easy to see and, in fact, straightforward to prove, that such a decimation strategy cannot lead to a numerically stable system with perfect reconstruction. Considering that many successful processing schemes for the STFT rely on both perfect reconstruction and the representation of the coefficients as a time-frequency matrix, this presents a notable obstruction to the adoption of wavelets in (audio) signal processing. 


\subsection{Quasi-Random Sequences}\label{ssec:quasirand}

Quasi-random sequences, also known as low-discrepancy sequences~\cite{Niederreiter1992random}, are deterministic sequences of numbers or $k$-dimensional coordinates that share some properties of uniformly distributed random numbers and can be used as a replacement for the latter in various applications, most prominently the quasi-Monte Carlo method for numerical integration~\cite{Niederreiter1992random}. The term \emph{discrepancy} refers to one of several related measures that quantify how uniformly distributed a set of points is. This property is important, e.g., to obtain error bounds in quasi-Monte Carlo integration. While most modern theoretical studies are concerned with quasi-random sequences of high dimensionality, we only require one-dimensional sequences. Specifically, any $N$ consecutive elements of a low-discrepancy sequence in dimension $D=1$ provides $N$ numbers that are particularly well distributed in the unit interval. 

Decimation strategies for wavelets can be derived from any low-discrepancy sequence. Here, we focus on two carefully chosen examples based on two prominent classes of quasi-random sequences. 

\vspace{6pt}
\noindent\textit{Kronecker sequences:} A deceptively simple construction are so-called Kronecker sequences which are of the form $(\{\alpha l\})_{l \geq 0} = (0,\{\alpha\},\{2\alpha\},\ldots)$ for some real $\alpha$, where $\{x\}:=x-\lfloor x \rfloor$ denotes the fractional part of a real $x$. It is well-known that Kronecker sequences are particularly well distributed if $\alpha$ is a \emph{badly approximable number}. These are irrational numbers that are particularly poorly approximated by rationals. Formally, a number $\alpha\in\RR$ is badly approximable if there is a constant $c>0$, such that
\[
  \left|\alpha - p/q\right| > c/q^2,
\]
for all nonzero integers $p,q \in \ZZ\setminus \{0\}$.  It is known that an irrational number is badly approximable if and only if the coefficients of its continued fraction expansion are bounded.
Among all the badly approximable numbers, the \emph{golden ratio} $\phi = \frac{1+\sqrt{5}}{2}$ maximizes the optimal constant $c$. Since this property is shared exactly with all so-called equivalent numbers of the form $\frac{a \phi +b}{c \phi + d}$, for integers $a,b,c,d$ with $ad-bc=\pm 1$, we may likewise consider the Kronecker sequence with $\alpha=1-1/\phi$, i.e., $a,c=1$, $b=-1$, and $d=0$.  For more information on Kronecker sequences we refer to the books \cite{DT97,kuinie}.

\vspace{6pt}
\noindent\textit{Digital $(0,1)$-sequences over $\ZZ_2$:} A $(0,1)$-sequence in base 2 is an infinite sequence $(x_l)_{l \geq 0} = (x_0,x_1,\ldots)$ in the unit-interval $[0,1)$ with the following property: for every $m \in \NN_0$ and every $k \in \{0,1,\ldots,2^m-1\}$, the elementary interval $[\frac{k}{2^m},\frac{k+1}{2^m})$  contains exactly one element of the point set 
\[\{x_l \ : \ p 2^m \leq l \leq (p+1) 2^m -1\}\quad \text{ for every } p \in \NN_0.\] 
One example for such a sequence is the well-known van der Corput sequence~\cite{vdc} (see also \cite{FAURE2015760}), which is related to the bit-reversal permutation. Typically, $(0,1)$-sequences in base 2 are constructed by the so-called digital method over the finite field $\ZZ_2$ of order two, i.e., the integers modulo $2$. To this end, let $\mathbf{C}=(c_{r,k})_{r,k \geq 1}$ be an $\infty \times \infty$-matrix over $\ZZ_2$, i.e., with entries $c_{r,k}$ from $\{0,1\}$, such that for every $m \in \NN$ the left-upper $m \times m$ sub-matrix is non-singular. Then the sequence $(x_l)_{l \geq 0}$ defined by 
\[x_l=\sum_{r=0}^{\infty}    \frac{\eta_{l,r}}{2^{r+1}} \quad \text{where }\ \eta_{l,r}=\sum_{k=0}^{\infty} c_{r+1,k+1} l_k \pmod{2},\]
where $l_k \in \{0,1\}$ are the binary digits of the index $l$, i.e., $l=l_0+l_1 2+l_2 2^2+\cdots$ (which are obviously 0 from a certain index on), is a $(0,1)$-sequence in base 2. 
In this context, the van der Corput sequence is covered by choosing the identity matrix for $\mathbf{C}$. We will use the specific digital $(0,1)$-sequence which is obtained from the infinite matrix 
\[
  \hspace{40pt} \mathbf{C} = \begin{pmatrix}
1 & 0 & 0 & 0 & 0 & \ldots \\
1 & 1 & 0 & 0 & 0 & \ldots \\
0 & 1 & 1 & 0 & 0 & \ldots \\
0 & 0 & 1 & 1 & 0 & \ldots \\
0 & 0 & 0 & 1 & 1 & \ldots \\
\vdots & \vdots & \vdots & \vdots& \vdots & \ddots
\end{pmatrix}\, .
 \]
For more information on (digital) $(0,1)$-sequences we refer to the books \cite{Niederreiter1992random,dick_pillichshammer_2010} or the survey article \cite[Section~3.2]{FAURE2015760}.

\section{A New Convention for Wavelet Sampling}\label{sec:oursampling}

Due to the geometric intuition outlined in Section \ref{ssec:wavelets}, linear spacing of wavelet center frequencies $\xi_j \sim s_j^{-1}$ has hardly been considered in the literature. For the same reason, uniform decimation, i.e., $d_j = d$ for some fixed $d>0$, is usually disregarded. We will now describe a decimation scheme that follows both of these conventions and which can be used to construct numerically stable, perfect reconstruction wavelet transforms with moderate, or even marginal, oversampling. \revised{By virtue of being a uniform filter bank frame, standard results~\cite{BoelcskeiFilterbanks1998,Janssen1998} ensure that the dual frame is a uniform filter bank as well, with the same decimation factor. }\revthree{The dual filter bank is, however, not necessarily a wavelet system. }
\revised{In particular, this implies that said dual frame can be precomputed using optimized factorization algorithms and, thus, synthesis from transform coefficients is highly efficient. 
The proposed scheme results in an \emph{almost} time-aligned decimated representation. 
More precisely, the coefficients across bands are in a position-independent, fixed relation, thus providing a meaningful notion of time frames and suggesting a natural arrangement in a time-frequency matrix.} 

Given a largest scale of interest $b>0$ (or equivalently a minimal frequency of interest $\xi_{\textrm{min}}>0$), a decimation factor $d>0$, a third parameter $q>0$ that determines the step size in the frequency direction, and finally a low-discrepancy sequence $(\delta_0,\delta_1,\delta_2,\ldots)$ determining channel specific delays, we select the translation-dilation pairs
\begin{equation}
  (x_{l,j},s_j) = \left(d(l+\delta_j),\frac{1}{b^{-1}+q^{-1}j}\right), 
\end{equation}
for all integers $l\in\ZZ$ and nonnegative integers $j\in\NN_0$. In other words, we consider the discrete wavelet system 
$\{\psi_{l,j}\}_{l\in\ZZ,j\in\NN_0}$, with 
\begin{equation}
   \psi_{l,j}(t) = \sqrt{\tfrac{1}{b}+\tfrac{j}{q}}\cdot\psi\left((\tfrac{1}{b}+\tfrac{j}{q})\cdot(t - d(l+\delta_j)\right).
\end{equation}

\begin{figure}[!t]
    \centering
    \includegraphics[width=0.5\textwidth,trim=9cm 5.1cm 4cm 0cm, clip]{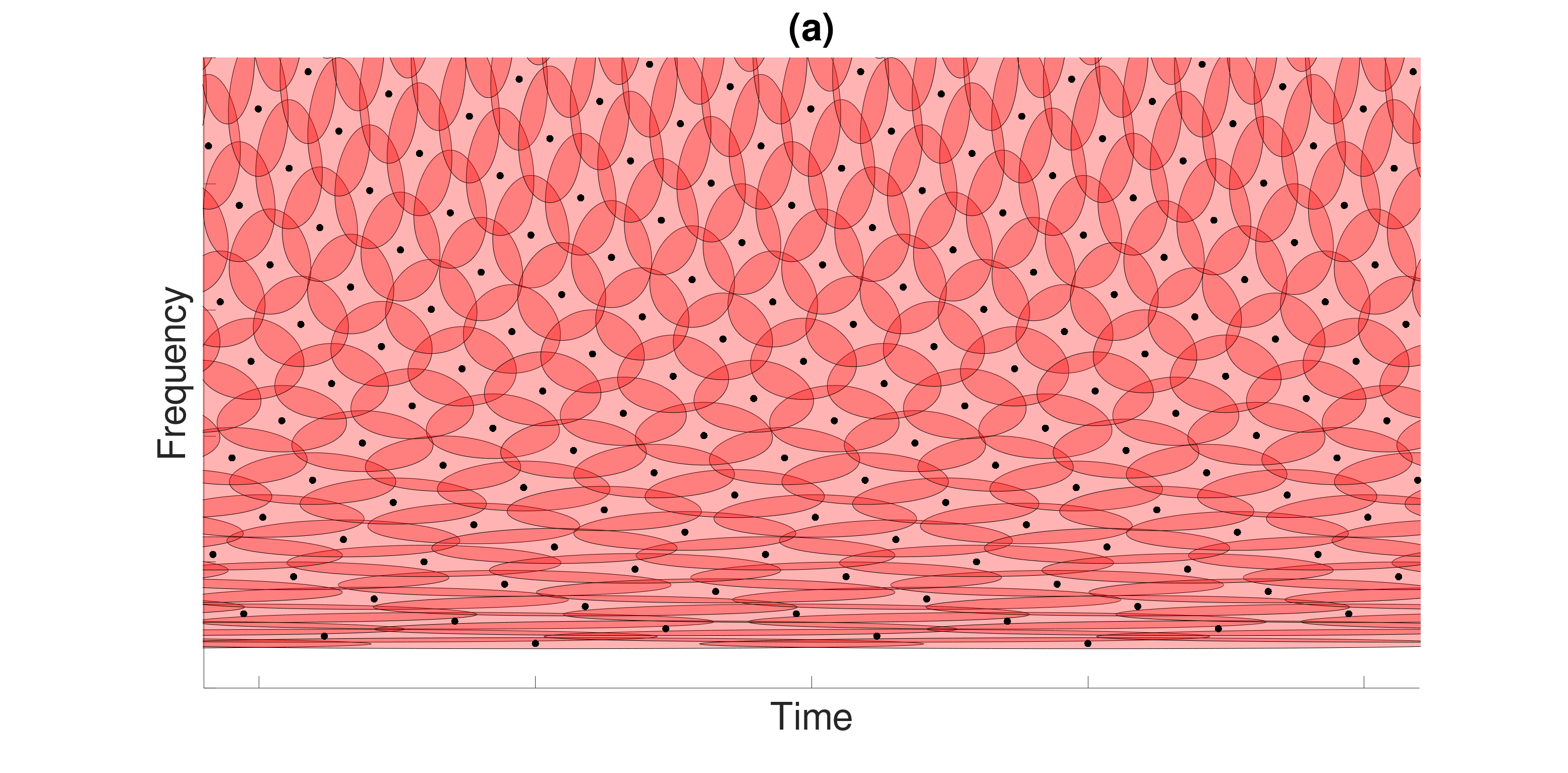}
    \includegraphics[width=0.5\textwidth,trim=9cm 2.1cm 4cm 0cm, clip]{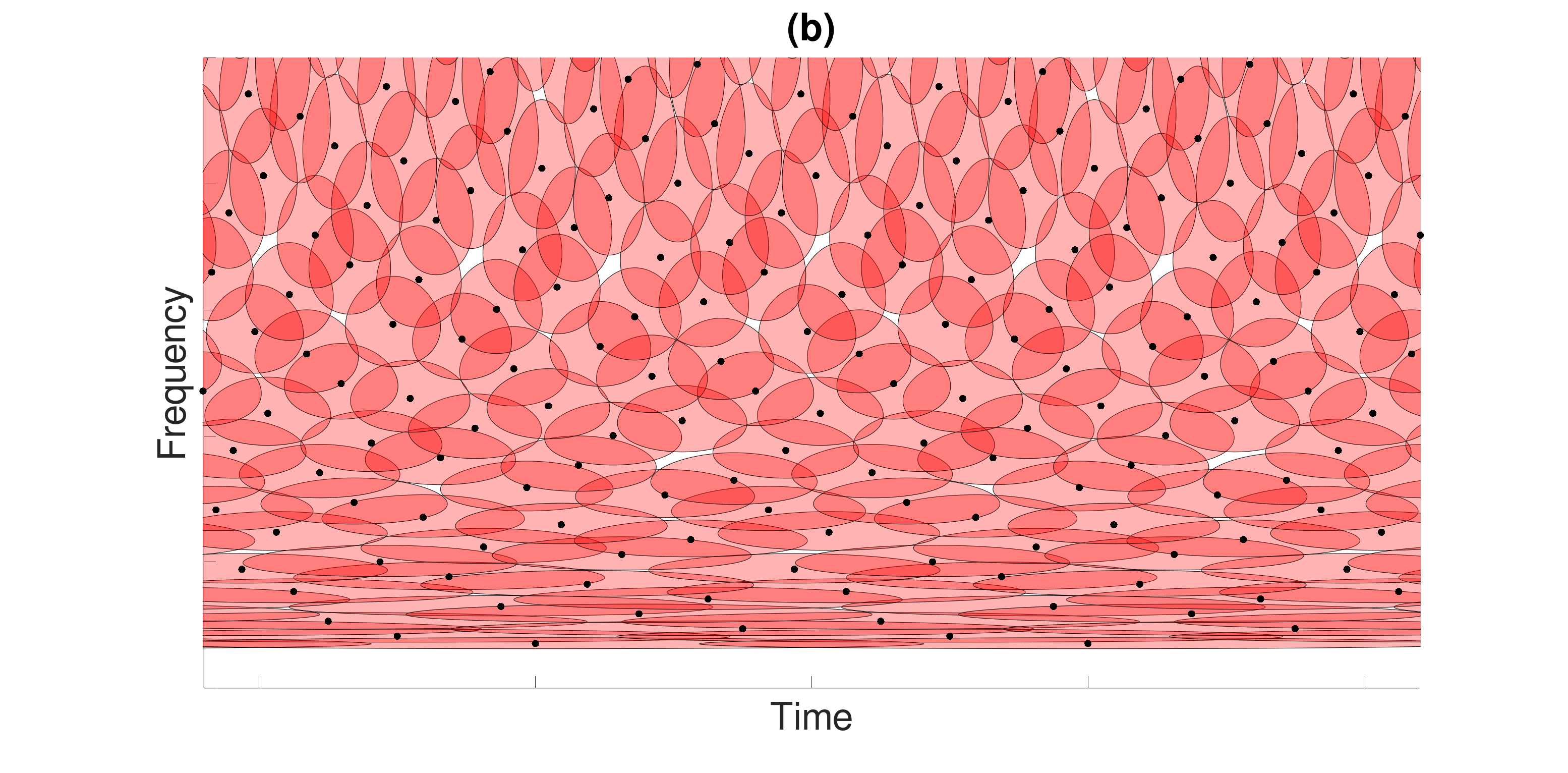}
\caption{Time-frequency geometry of decimated wavelet systems with quasi-random delay: (a) Decimation on a uniform grid with delays chosen according to the \emph{golden} Kronecker sequence provides an even covering comparable to the classic wavelet decimation (Fig.~\ref{fig:regularwavelets}(a)). (b) Deriving the delays instead from the digital sequence described in Section \ref{ssec:quasirand}, we obtain a covering that is slightly less even. In both cases, the resulting covering will become uneven below a certain frequency, similar to the bottom region in Fig.~\ref{fig:regularwavelets}(b).
 \revisedtwo{Note that the radius of the depicted hyperbolic circles is chosen as in Fig.~\ref{fig:regularwavelets} to allow comparison. A mild increase in radius is sufficient to remove the \emph{blind spots} in panel (b), whereas even a minor decrease would introduce similar blind spots in panel (a) and Fig.~\ref{fig:regularwavelets}(a).}
\label{fig:quasi-randomwavelets}}
\end{figure}

In contrast to the rectangular grid shown in Fig.~\ref{fig:regularwavelets}(b), which can be generated by the above construction with delays $\delta_j=0$ for all $j$, choosing the $\delta_j$ as the elements of a quasi-random sequence provides a wavelet system that covers the time-frequency plane surprisingly evenly, see Fig.~\ref{fig:quasi-randomwavelets}. Intriguingly, if we choose the $\delta_j$ as the elements of a Kronecker sequence, the points $\{(x_{l,j},\xi_j)\}_{l\in\ZZ,j\in\NN_0}$ form a uniform (skewed) time-frequency grid. 

\subsection{Numerically Stable Wavelet Transforms with Perfect Reconstruction and Uniform Decimation}\label{ssec:perfectreconstruction}

\begin{figure}[!t]
    \centering
    \includegraphics[width=0.5\textwidth,trim=4.7cm 1.4cm 1.2cm 0.1cm, clip]{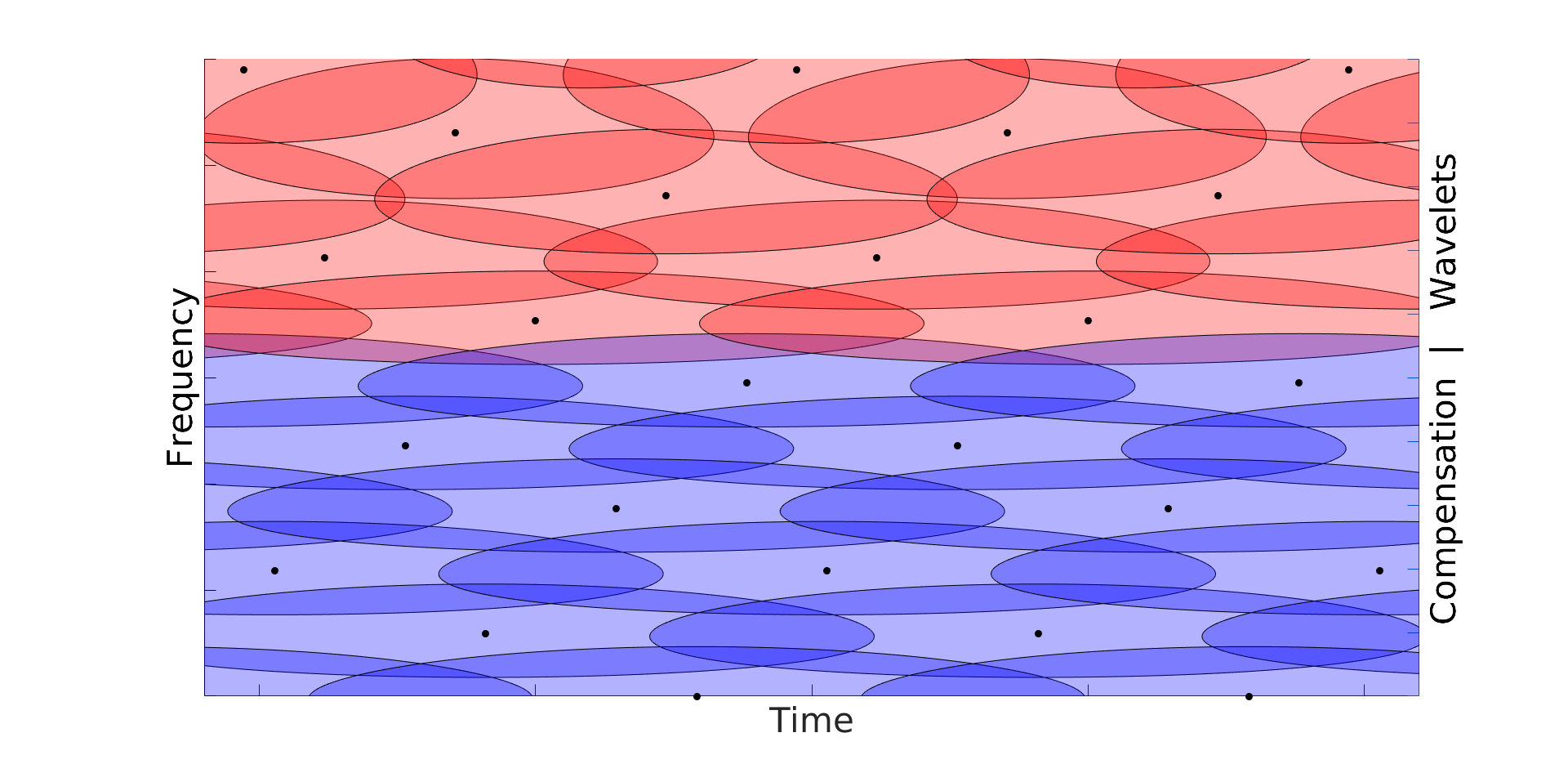}
\caption{\revised{Proposed covering of the low-frequency region by  modulated and delayed versions of the largest scale wavelet filters for the decimated wavelet system with
quasi-random delays chosen
according to the golden Kronecker sequence.
The frequency axis is rescaled compared to Fig.~\ref{fig:quasi-randomwavelets}(a) for better visibility. }\label{fig:quasi-randomwavelets-lf}}
\end{figure}

As indicated by Fig.~\ref{fig:quasi-randomwavelets}, the proposed wavelet decimation yields a surprisingly uniform covering of the time-frequency plane. However, due to the uniform spacing in frequency, this is only true as long as the scale of the wavelets is small enough (relative to the frequency step parameter $q$). Since the wavelet bandwidth is proportional to its center frequency, we find that the lower frequency region is insufficiently covered, similar to Fig. \ref{fig:regularwavelets}(b), for arbitrarily large base scale $b$. The size of this region depends on the chosen wavelet and the frequency step parameter $q$. To compensate for this lack of coverage, we introduce additional 
\revised{\emph{compensation filters} covering the low-frequency region}
\begin{equation}\label{eq:lowpassfilters}
   \psi_{l,j}(t) = \frac{1}{\sqrt{b}}\psi\left(\frac{t - d(l+\delta_j)}{b}\right)e^{2\pi i \xi_1 \cdot \tfrac{j(t-d(l+\delta_j))}{q}},
\end{equation}
for negative integers $j\geq -q/b$. 
In words, we demodulate the wavelet at the base scale $b$ in uniform steps for as long as the center frequency remains nonnegative. Note that the construction above implies that the index of the quasi-random sequence of delays is shifted accordingly, i.e., we denote the $j$-th element by $\delta_{j-\lfloor q/b\rfloor}$ instead of $\delta_j$.
\revised{This construction is illustrated in Fig.~\ref{fig:quasi-randomwavelets-lf}, where we rescaled the frequency axis compared to Fig.~\ref{fig:quasi-randomwavelets}(a) (but not the time axis) for better visibility.}
Clearly, this is only one of many possibilities for the construction of suitable 
\revised{filters covering the low-frequency region}, chosen here because it preserves the filter prototype, or mother wavelet, and the uniform decimation of the system in time and frequency. 

The full system $\{\psi_{l,j}\}_{l\in\ZZ,j\in\ZZ\cap [-q/b,\infty)}$ remains uniformly decimated and its properties, in particular the perfect reconstruction property and numerical stability can be studied using frame theory for uniform filter banks~\cite{BoelcskeiFilterbanks1998} or, equivalently, shift-invariant systems~\cite{Janssen1998}, which also provides highly efficient algorithms for reconstruction from the filter bank coefficients. 
We observe that, in practice, choosing the decimation parameters $d$ and $1/q$ small enough is sufficient to obtain an invertible, numerically stable filter bank. In Section~\ref{ssec:numeval}, we validate this assessment \revisedtwo{for finite wavelet systems in $\CC^L$}.

\revised{
\section{Numerical Evaluation, Complexity, and Implementation}\label{sec:computation}

Although the parametrization of $\{\psi_{l,j}\}_{l,j}$ introduced in the previous section in terms of $b>0$ and $q>0$ arises naturally, 
it is inconvenient for practical construction. Instead, we subsequently adopt the following parametrization, where we restrict to the case $q/b\in \NN$. 
We select a mother wavelet $\psi$, the desired number $M+1$ of frequency channels equidistantly spaced at center frequencies in $[0,\xi_\textrm{samp}/2]$, where $\xi_\textrm{samp}$ is the sampling rate, and the number $M_{\textrm{C}}$ of compensation filters. 
The relation to the parametrization introduced above is given by $M_{\textrm{C}} = q/b$ and $\xi_1/q = \xi_\textrm{samp}/(2M)$.
In the provided implementation, the wavelet is automatically scaled such that the center frequency of $\psi_{l,0}$ equals $\xi_\textrm{samp}\cdot M_{\textrm{C}}/(2M)$. The redundancy of the resulting system is then controlled by selecting the decimation factor $d$, similar to a common approach for parametrizing the STFT. 
}

\subsection{Numerical Evaluation}\label{ssec:numeval}
To validate that the proposed decimation strategy indeed leads to invertible, numerically stable wavelet systems, we compute the \revised{frame bound ratio $R_\textrm{FB}$ for various wavelet filter banks so decimated. Here, we consider filter banks acting on finite sequences, in which case, this ratio equals the condition number of the matrix that realizes the composition of filter bank analysis and synthesis and can be computed directly. To do so, we rely on the \emph{Large Time-Frequency Analysis Toolbox} (LTFAT, \url{ltfat.org}), which achieves this computation by means of an efficient factorization of said matrix. Recall that the filter bank is invertible if and only if this $R_\textrm{FB}$ is finite and stable if it is small, with perfect stability if $R_\textrm{FB}=1$.}
The evaluation presented here can be reproduced using the code available at \url{ltfat.org/notes/057}.

In a pre-test for the proposed evaluation, we noticed that for fixed $M$, the frame bound ratio $R_\textrm{FB}$ decreases monotonically with increasing $M_{\textrm{C}}$, up to a certain value of $M_{\textrm{C}}$ after which a further increase yields no benefit. Furthermore, we found that this value of $M_{\textrm{C}}$ does not depend on the choice of $M$. Increasing $M_{\textrm{C}}$ stabilizes the frequency response of the filter bank, which equals $\Psi = \sum_{j} |\widehat{\psi_{0,j}}|^2$ up to a positive multiplicative constant, especially in the low-frequency region. Strong fluctuations of $\Psi$ have detrimental effect on $R_\textrm{FB}$, and increasing $M_{\textrm{C}}$ reduces these fluctuation, explaining the first observed effect. The lack of a dependence on $M$ can be explained by the fact that the frequency response of two such filter banks that only differ in the choice of $M$ are equal up to a dilation and a positive multiplicative constant. Altogether, these observations allow us to greatly reduce the number of tested configurations. 

\begin{table*}[!t]
\begin{center}
\caption{Optimized frame bound ratios for the proposed wavelet systems. Table values are $R_\textrm{FB}$ ($M_{\textrm{C}}$, $M$), i.e., the optimal frame bound ratio $R_\textrm{FB}$ is achieved with $M_{\textrm{C}}$ \revised{compensation } channels and $M$ total channels.}\label{tab:framebounds}
\begin{tabular}{cccccccc}
  \toprule
  \multicolumn{8}{c}{Kronecker-sequence delays}\\
  \midrule
  & \multicolumn{4}{c}{Cauchy} & \multicolumn{3}{c}{B-Spline}\\
  \cmidrule(lr){2-5} \cmidrule(lr){6-8}
  Oversampling & $\alpha = 100$ & $\alpha = 300$ & $\alpha = 900$ & $\alpha = 2700$ & $\xi_{\textrm{fm}} = 3$ & $\xi_{\textrm{fm}} = 6$ & $\xi_{\textrm{fm}} = 10$\\
  \midrule
  1.2 &  $15.06$ $(2,102)$&   $14.17$ $(4,202)$ & $13.61$ $( 6,307)$ & $13.74$ $(11, 550)$& $15.04$ $(4,202)$ & $14.08$ $( 7,449)$ & $14.08$ $(12, 620)$\\
     2 & $ 3.22$ $(3,448)$  &  $ 2.98$ $(5,253)$ & $ 2.92$ $( 9,501)$ & $ 2.94$ $(15, 764)$& $ 3.01$ $(5,253)$ & $ 2.93$ $( 9,511)$ & $ 2.97$ $(15, 764)$\\
     4 & $ 1.72$ $(4,768)$  &  $ 1.62$ $(7,350)$ & $ 1.60$ $(12,768)$ & $ 1.59$ $(20,1012)$& $ 1.60$ $(6,363)$ & $ 1.59$ $(12,768)$ & $ 1.59$ $(20,1024)$\\
     8 & $ 1.31$ $(4,214)$  &  $ 1.25$ $(8,404)$ & $ 1.23$ $(14,702)$ & $ 1.24$ $(26,1306)$& $ 1.25$ $(7,473)$ & $ 1.24$ $(15,757)$ & $ 1.25$ $(25,1250)$\\
  \midrule
  \multicolumn{8}{c}{Digital (0,1)-sequence delays}\\
  \midrule
  & \multicolumn{4}{c}{Cauchy} & \multicolumn{3}{c}{B-Spline}\\
  \cmidrule(lr){2-5} \cmidrule(lr){6-8}
  Oversampling & $\alpha = 100$ & $\alpha = 300$ & $\alpha = 900$ & $\alpha = 2700$ & $\xi_{\textrm{fm}} = 3$ & $\xi_{\textrm{fm}} = 6$ & $\xi_{\textrm{fm}} = 10$\\
  \midrule
  1.2 & $20.38 $ $(2, 127)$&    $19.27 $ $(4, 255)$ & $17.48  $ $(6,  384)$ & $16.02  $ $(11, 1023) $& $20.67 $ $(4, 256)$ & $17.26 $ $(7,  384) $ & $16.98 $ $(12, 640) $\\
   2 & $3.80 $ $(3, 260)$  &  $3.73 $ $(5, 384) $ & $3.66  $ $(8,  656) $ & $ 3.57 $ $(14,  771)  $& $ 3.72 $ $(5, 388)$ & $ 3.62  $ $(9,  771)$ & $ 3.62 $ $(14, 771) $\\
   4 & $1.74 $ $(4, 255)$  &  $1.67 $ $(7, 383) $ & $1.63 $ $(12,  639) $ & $ 1.62 $ $(21, 1279)  $& $ 1.78 $ $(6, 640)$ & $ 1.62 $ $(12,  639)$ & $ 1.60 $ $(20, 1023)$\\
   8 & $1.27 $ $(5, 256)$  &  $1.20 $ $(9, 511) $ & $1.20 $ $(16, 1535) $ & $ 1.21 $ $(27, 1791)  $& $ 1.21 $ $(8, 511)$ & $ 1.27 $ $(16, 1408)$ & $ 1.22 $ $(27, 1791)$\\
  \bottomrule
\end{tabular}
 \end{center}
\end{table*}

With the above considerations in mind, we perform an optimization of the frame bounds by first determining, for a fixed mother wavelet and oversampling rate, the smallest value of $M_{\textrm{C}}$ that yields optimal $R_\textrm{FB}$. We then proceed to determine the value $M\in\{128,\allowbreak 256,\allowbreak 384,\allowbreak 512,\allowbreak 640,\allowbreak 768,\allowbreak 1024,\allowbreak 1280,\allowbreak 1536,\allowbreak 2048\}$ that optimizes $R_\textrm{FB}$, with the additional restriction that $M\geq 50 M_{\textrm{C}}$, by an exhaustive search. Finally, we further refine the choice of $M$ by a divide and conquer approach starting from the determined optimizer and its two neighbors in the sequence of tested values for $M$. The restriction $M_{\textrm{C}}\leq M/50$ ensures that the center frequency of $\psi_{l,
0}$ is no larger than $\xi_\textrm{samp}/100$, e.g., $441$\,Hz for a sampling rate of $44.1$\,kHz. 

\vspace{6pt}
\noindent\textit{Results: }
In Table \ref{tab:framebounds}, we show the obtained optimal frame bounds for oversampling rates $1.2,\ 2,\ 4$, and $8$. As mother wavelet, we consider Cauchy wavelets~\cite{dapa88}, defined (up to a normalization constant) by $\widehat{\psi}(\xi) = \xi^{\frac{\alpha-1}{2}} e^{- 2\pi \xi}$, for $\alpha$ equal to $100,\ 300,\ 900$, and $2700$, where higher $\alpha$ implies a higher Q-factor.%
\footnote{\revised{The Q-factors for these hyperparameters have been estimated as the ratio of the mother wavelets' center frequency 
and their bandwidth at $-3 \text{dB}$ height relative to their maximum. They correspond to $2.9999$, $5.2053$, $9.0212$, and $15.6282$, respectively.}}
As an example of a compactly supported wavelet, we consider the modulated fourth order B-spline defined (up to a normalization constant) by $\widehat{\psi}(\xi) = \sin(\pi(\xi-\xi_{\textrm{fm}}))^4/(\pi(\xi-\xi_{\textrm{fm}}))^4$, with $\xi_{\textrm{fm}}$ equal to $3,\ 6$, and $10$, implying Q-factors that are roughly equivalent to Cauchy wavelets with $\alpha$ equal to $257$, $1024$, and $2842$, respectively\footnote{\revised{The estimated Q-factors for the B-spline wavelets correspond to $4.1600$, $8.3200$, and $13.8666$, while those for the Cauchy wavelets correspond to $4.8171$, $9.6229$, and $16.0340$, respectively.}}, see \cite{ltfatnote055}. 

As is to be expected, the optimal frame bound ratio decreases with higher oversampling rate. For fixed oversampling, however, there is only limited dependence on the mother wavelet and Q-factor. 
Specifically, we obtain $R_\textrm{FB}$ close to $14,\ 3,\ 1.65$, and $1.3$ for oversampling rates $1.2,\ 2,\ 4$, and $8$ across almost all conditions with Kronecker-sequence based decimation. 
For redundancy $4$ and below, the results for the decimation based on a digital-(0,1) sequence yield slightly worse stability, i.e., larger $R_\textrm{FB}$. 
Interestingly, for factor $8$ oversampling, $R_\textrm{FB}$ is improved by using this decimation scheme, for all but one condition, indicating that a decimation scheme based on digital-(0,1) sequences is beneficial when large oversampling rates are used. 

\begin{figure}[!t]
\includegraphics[width=0.49\textwidth,height=.4\textwidth,trim=0cm 0cm 0.75cm 0.75cm, clip]{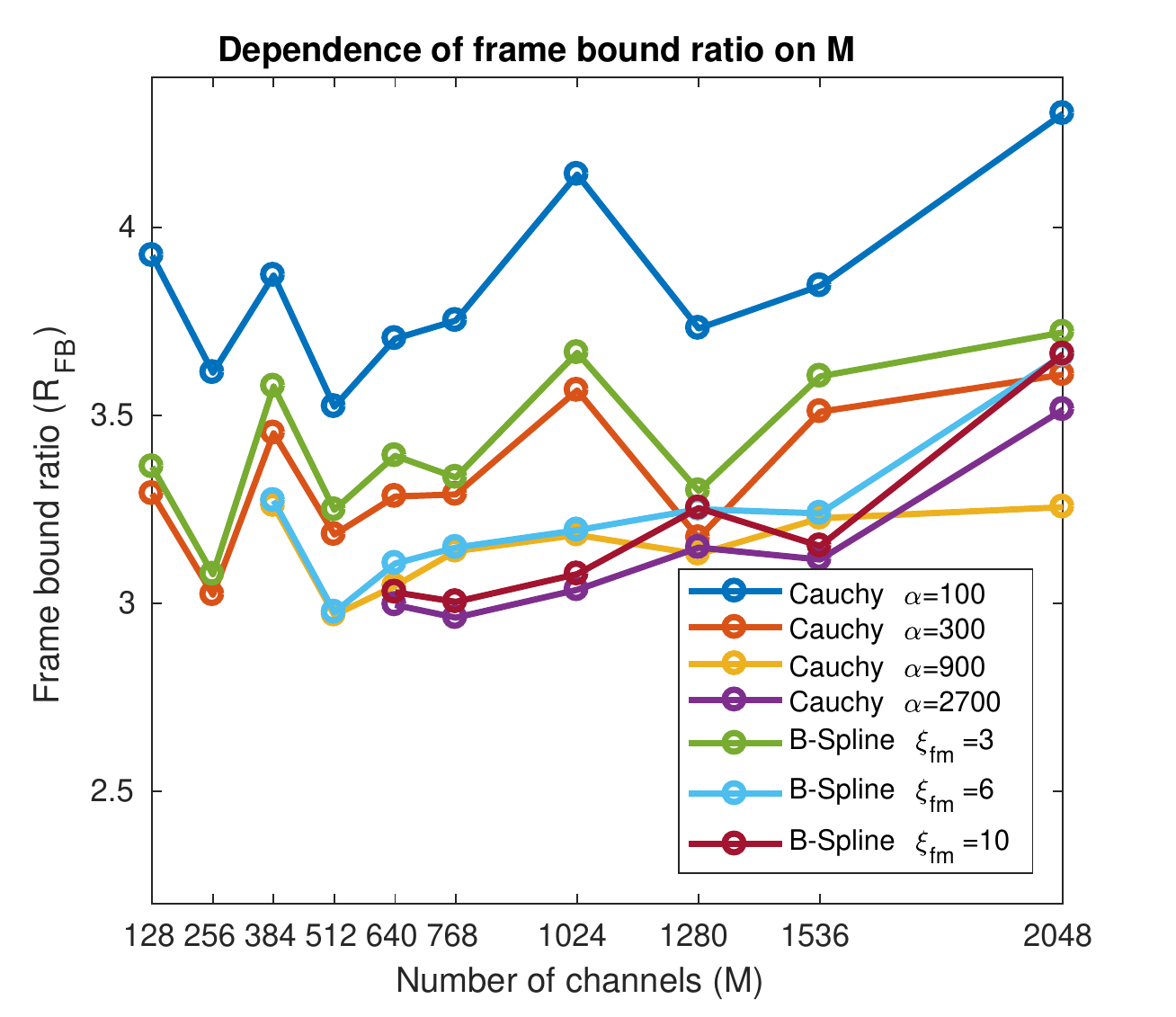}
\caption{Examples for the dependence of $R_{\textrm{FB}}$ on $M$ for Kronecker-sequence delays and oversampling rate $2$.  \label{fig:BoundExperiment}}
\end{figure}

Note that our divide and conquer refinement procedure is only guaranteed to find the global minimum of $R_\textrm{FB}$ as a function in $M$ if this function is convex. 
This is usually not the case as illustrated in Fig.~\ref{fig:BoundExperiment}, where we \revised{plot the values of $R_\textrm{FB}$ at oversampling rate $2$, for all choices of $M$ considered in the exhaustive search described above and all seven considered mother wavelets.} 
Thus, it may still be possible to improve the reported bounds. 
Furthermore, this is an explanation why the reported optimal parameter $M$ does not depend monotonically on the Q-factor of the mother wavelet which is larger for larger choices of $\alpha$ or $\xi_{\textrm{fm}}$.

\begin{figure*}[!t]
\includegraphics[width=0.244\textwidth,trim=2.5cm 0.5cm 8.2cm 0cm, clip]{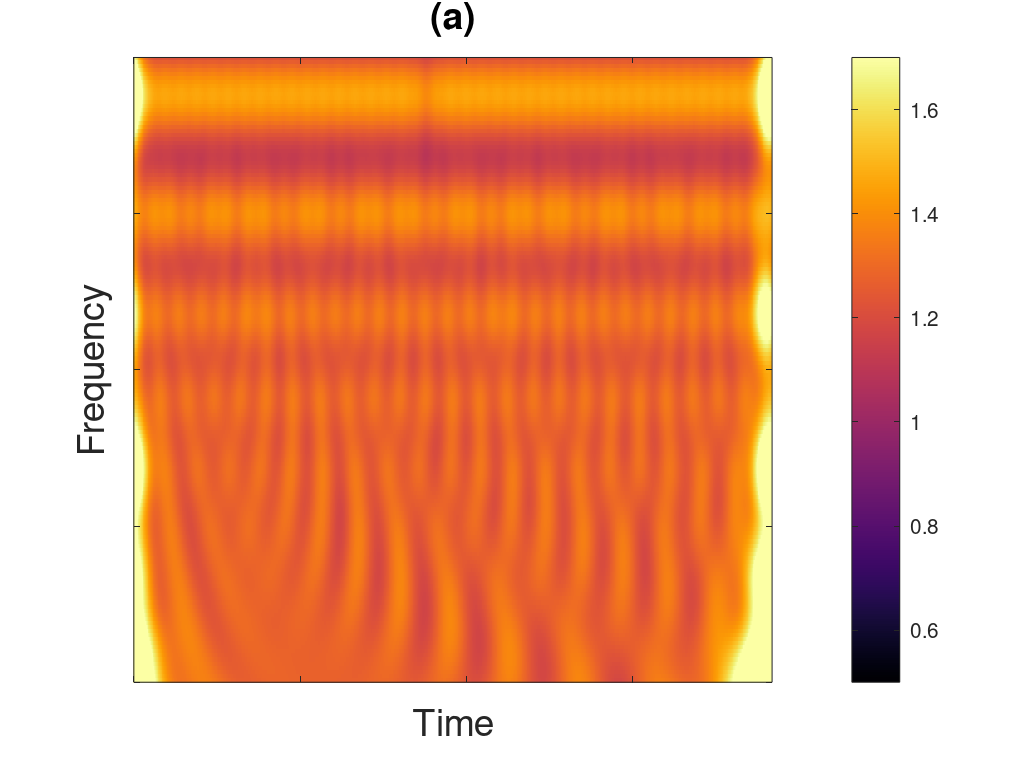}
\includegraphics[width=0.225\textwidth,trim=4.5cm 0.5cm 8.2cm 0cm, clip]{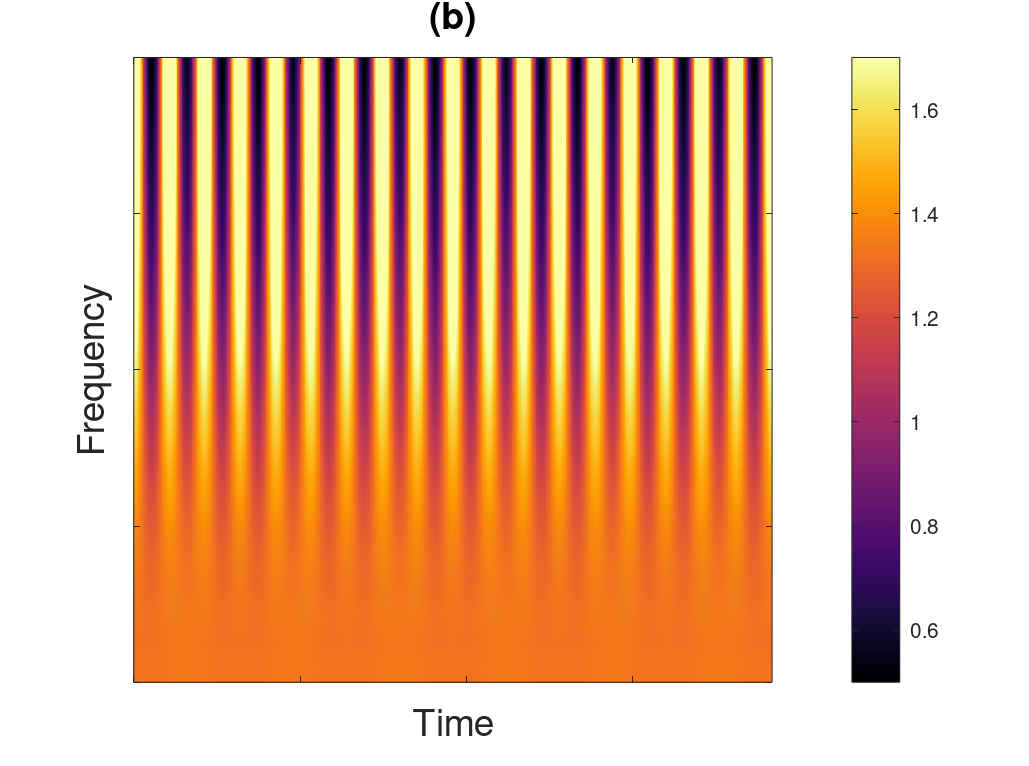}
\includegraphics[width=0.225\textwidth,trim=4.5cm 0.5cm 8.2cm 0cm, clip]{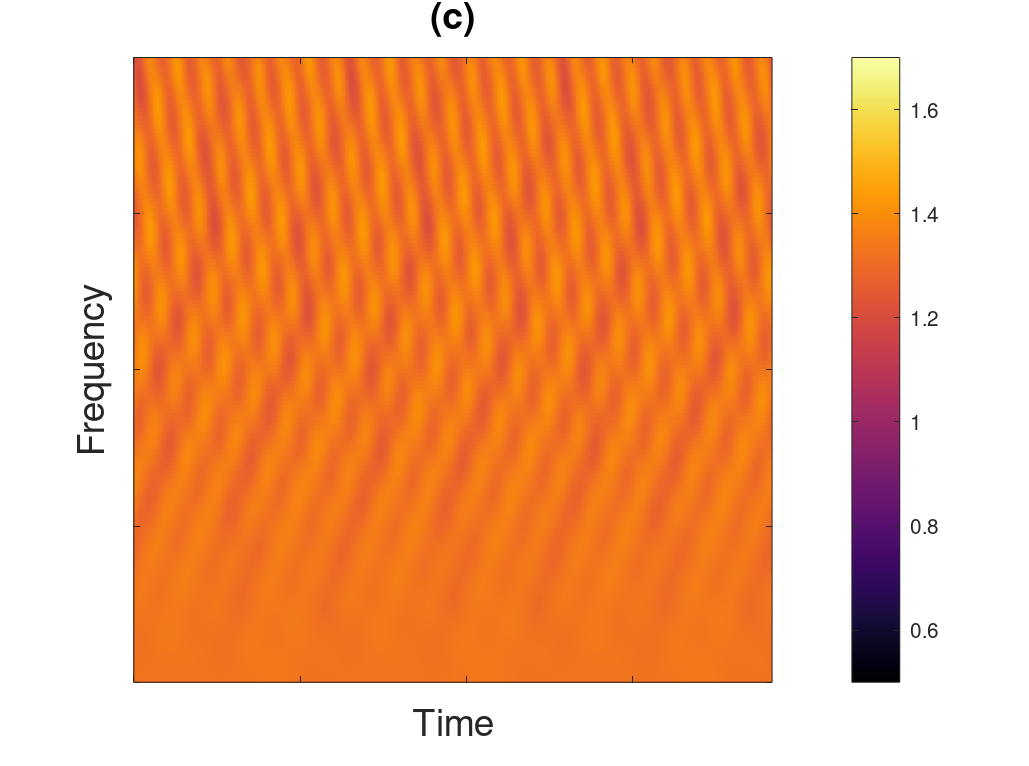}
\includegraphics[width=0.291\textwidth,trim=4.5cm 0.5cm 1.2cm 0cm, clip]{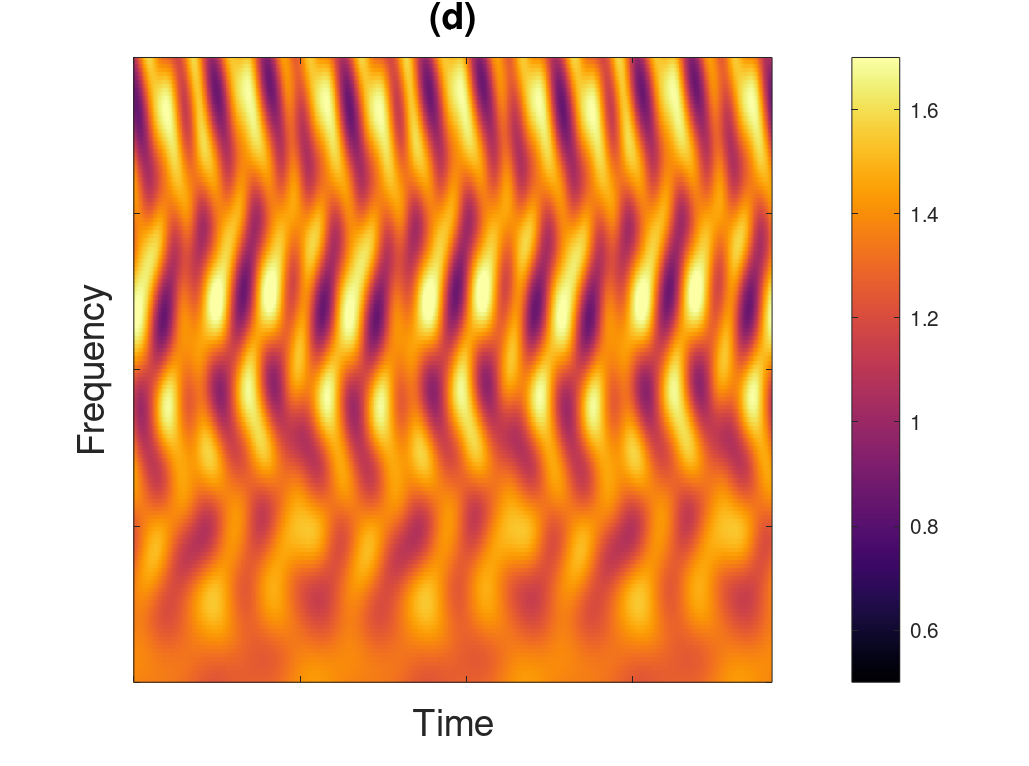}
\caption{Accumulated spectrograms of decimated wavelet systems: (a) classical nonuniform decimation, (b) uniform grid, (c) uniform grid with delays chosen according to the \emph{golden} Kronecker sequence, (d) delays derived from the digital sequence described in Section \ref{ssec:quasirand}.\label{fig:accuspec}}
\end{figure*}

To validate that the wavelet coverings illustrated in Figs.~\ref{fig:regularwavelets} and \ref{fig:quasi-randomwavelets} above do indeed conform to the time-frequency domain being well covered by the proposed wavelet systems, we calculated accumulated spectrograms~\cite{abreuAccuSpecs} of these systems \revised{at an oversampling rate of approximately $2$, see Fig.~\ref{fig:accuspec}.}
More specifically, we calculated the spectrogram for each wavelet in the system using a short-time Fourier transform with a Gaussian window and summed all of them.
Since the spectrogram of a signal is a representation of its time-frequency energy localization, this sum illustrates the time-frequency area that is well represented by the different wavelet systems. 
Our simulations 
essentially confirm the findings of Fig.~\ref{fig:regularwavelets} and \ref{fig:quasi-randomwavelets}.
The classical wavelet system in Fig.~\ref{fig:accuspec}(a) and the delay shifted system using the \emph{golden} Kronecker sequence in Fig.~\ref{fig:accuspec}(c) show the best uniformity. Clearly, the uniform system without delays in Fig.~\ref{fig:accuspec}(b) overemphasizes certain time-frequency regions while completely missing the area in between. 
The uniform system with delays based on a digital $(0,1)$-sequence in Fig.~\ref{fig:accuspec}(d) does not cover the area quite as uniformly. 
However, as the accumulated spectrograms presented here were obtained with two-fold oversampling, 
this finding matches the computed frame bounds: 
%
At low oversampling rates, the Kronecker-sequence delays are superior to the digital $(0,1)$-sequence delays.

\subsection{Block-Processing and Complexity}\label{ssec:complexity}
\revised{
The most straightforward, efficient implementation of the proposed wavelet decimation implements the forward and the backward transform as FFT filter banks~\cite{smith2009audio}, similar to  \cite{velasco2011constructing,schoerkhuber2014a}. Compared to these references, this type of implementation of the proposed decimation implies a moderately higher computational load, due to its large number of overlapping frequency channels. By construction, FFT filter banks process the entire input signal at once. Nonetheless, it is possible to achieve block-wise processing by adopting the \emph{slicing} scheme proposed in \cite{dogrhove13}. Sliced processing requires the segmentation of an incoming data stream into large blocks and the incurred delay may not be acceptable for certain applications. The implementation of wavelet filter banks with short block-length is often quite intricate, not least due to their usually non-uniform decimation, see, e.g.,~\cite{ltfatnote026}. In our setting, however, shorter block-length, and thus reduced delay, can be achieved with a straightforward time-domain implementation, using a mother wavelet with finite support. Although real-time implementation is not our main objective, we outline the computational cost of such an implementation.

The proposed uniform decimation admits an implementation with fixed block length, where each block corresponds to a time frame that contains one filter bank coefficient per channel.
The relative temporal positions of all coefficients are identical in each time frame. 
Such an implementation is computationally straightforward and achieved by computing the inner products between the input signal and the wavelet atoms directly, see \eqref{eq:waveletttransform}. Due to the constant Q-factor of the wavelet transform, the computational cost of doing so depends only logarithmically on the number of channels. 
Specifically, we compute the wavelet coefficients with a decimation factor $d$ for $M+1$ equidistant frequency channels, the first $M_\textrm{C}$ of which are compensation filters. We assume that, at the largest considered scale, the wavelet impulse response has a length of $L_\textrm{W}$ samples. Since the delays are in the interval $[0,d)$, we can segment the input into time frames, or blocks, of length $L_\textrm{B} = L_\textrm{W} + d$, with an overlap of $L_\textrm{W}$ samples. In each time frame, we compute one coefficient per channel. 
The total cost per time frame for computing the compensation filters is $M_{\textrm{C}}\cdot L_\textrm{W}$ multiplications and additions each. For the remaining filters, the impulse response length decreases as scales decrease and one can easily see that  
\begin{equation}\label{eq:cplxity1}
  \begin{split}
  L_\textrm{W}\cdot \sum_{j=M_{\textrm{C}}}^M \frac{M_{\textrm{C}}}{j} & \leq M_{\textrm{C}}L_\textrm{W} \cdot \int_{M_{\textrm{C}}-1}^M s^{-1}~ds \\
  & = M_{\textrm{C}} L_\textrm{W}\cdot \ln(M/(M_{\textrm{C}}-1))
  \end{split}
\end{equation}
multiplications and additions are required per time frame.
Here, we used the standard integral estimate for partial sums of harmonic series. Note that the above computation ignores the rounding of impulse response lengths to the next integer and assumes that $M_{\textrm{C}}\geq 2$. Overall, a direct implementation of the forward transform amounts to approximately  
\begin{equation}\label{eq:cplxity2}
  M_{\textrm{C}} L_\textrm{W}\cdot (1+\ln(M/(M_{\textrm{C}}-1))
\end{equation}
multiplications and additions per time frame and introduces a  delay of $L_\textrm{B}$ samples. A more sophisticated implementation employing low-pass filtering and subsequent decimation in the spirit of \cite{schorkhuber2010constant} can introduce significant optimization at the cost of a small error in the coefficient computation. 

Computation of the backwards transform with perfect reconstruction relies on the dual filters (see Section \ref{ssec:wavelets}).
For the wavelet configurations in the present work, the essential impulse response length of the dual filters is of the same order as that of the corresponding original filter. This implies that  the backward transform can be achieved at similar computational cost and delay as the forward transform. However, since the dual impulse responses are not expected to have finite support, they must be approximated, introducing a trade-off between accuracy and delay. }

\subsection{Wavelets in the Large Time-Frequency Analysis Toolbox}
\label{sec:ltfat}

\revised{All computations in Sections \ref{ssec:numeval} and \ref{sec:experiments} rely on the implementation of wavelet filter banks in the Large Time-Frequency Analysis Toolbox (LTFAT), updated to support the proposed wavelet decimation with the release of LTFAT 2.5.0.
In particular, the function \texttt{waveletfilters} supplies the frequency responses of the wavelet filters, as specified by the given input parameters, alongside a set of decimation factors and the filters' center frequencies. By default, \texttt{waveletfilters} accepts the filter bank length $L$ and a vector of wavelet scales, where the  unit scale $s=1$ corresponds to a center frequency of  $\xi_\textrm{samp}/20$, as input arguments. Additionally, a number of increasingly specific, optional input parameters can be used to customize the wavelet filters and cover a large number of use cases, including those in~\cite{ltfatnote053,ltfatnote055} and in the present work. As an alternative to providing the wavelet scales directly, \texttt{waveletfilters} supports the automatic allocation of filters in a specified frequency range, spaced either geometrically with a set number of bins per octave, or linearly with a set number of channels.

Most important for the present work, \texttt{waveletfilters} provides options for controlling the mother wavelet and decimation settings, as well as the oversampling rate. }By default, \texttt{waveletfilters} 
selects a Cauchy-type mother wavelet with $\alpha=300$ and non-uniform, integer decimation factors adapted to the wavelet bandwidth. 
Currently, the Cauchy~\cite{daubechies88}, Morse~\cite{olwa02}, Morlet, frequency B-spline~\cite{gao2010wavelets}, analytic spline~\cite{chaudhuryunser09}, and complex spline wavelets are implemented for positive and negative scales. The individual filters are generated by the separate function \texttt{freqwavelet}, enabling the future addition of further wavelet prototypes. \texttt{waveletfilters} \revised{supports uniform decimation and non-uniform decimation, with integer or fractional decimation factors. To construct perfect reconstruction filter banks, \texttt{waveletfilters} supplies two options for covering the frequency range from 0 Hz to the center frequency of the largest wavelet scale: By a single low-pass filter, or by several frequency-shifted copies of the filter corresponding to the largest wavelet scale, as described in Section~\ref{ssec:perfectreconstruction}.}

\revised{
For example, the call \texttt{\justify [g,a,fc]=waveletfilters( Ls,'linear',2,MC/M,1,M\textminus MC+1,\{'cauchy',900\}, 'uniform','redtar',4,'repeat','delay',@dly)} supplies a wavelet filter bank for signals of length $\texttt{Ls}$ with linear frequency spacing. Considering a sampling rate of $2$,  \texttt{M\textminus MC+1} wavelet scales are spaced between frequencies \texttt{MC/M} and 1 (Nyquist). The filter bank uses a Cauchy wavelet with $\alpha=900$, uniform decimation and a target oversampling rate of $4$. Finally, the flag $\texttt{repeat}$ adds \texttt{MC} compensation filters between frequency $0$ and \texttt{MC/M}, resulting in \texttt{M+1} total channels. The function \texttt{@dly} is used to generate the desired sequence of delays, and will be called internally, with the number of required sequence elements \texttt{M+1} and the vector of decimation factors \texttt{a} as input arguments. The output \texttt{[g,a,fc]} comprises a cell array \texttt{g} of filter frequency responses, a vector \texttt{a} of decimation factors and a vector \texttt{fc} of center frequencies.

The filterbank module of LTFAT provides a host of functions for working with and analyzing a filter bank so created. The function \texttt{filterbank} calculates the filter bank coefficients via fast FFT-based convolution. It accepts as input a target signal, as well as the wavelet filters and their associated decimation factors, as provided by \texttt{waveletfilters}.
 Synthesis from filter bank coefficients is realized by \texttt{ifilterbank} and their visualization can be achieved via \texttt{plotfilterbank}. If analysis-synthesis filter bank pairs with perfect reconstruction are desired, a dual filter bank can be obtained by applying  \texttt{filterbankdual} to the output of \texttt{waveletfilters}, respectively \texttt{filterbankrealdual} for filter banks that cover only positive frequencies. Note that this is only possible for uniform filter banks or under strict conditions on non-uniform filter banks. Generally, perfect reconstruction can be achieved with an iterative method, implemented in \texttt{ifilterbankiter}, provided that the analysis filter bank is invertible. The frame bounds of a filter bank can be obtained by \texttt{filterbankbounds} and \texttt{filterbankrealbounds} respectively, e.g., to verify invertibility. The LTFAT filterbank module provides selected methods for advanced functionality, such as phaseless reconstruction~\cite{ltfatnote021,ltfatnote051} and time-frequency reassignment~\cite{ltfatnote041}.
 
 Finally, the output of \texttt{waveletfilters} is compatible with the block processing framework in LTFAT, which enables experimental real-time application by implementing a variant of the sliced processing proposed in~\cite{dogrhove13}. 
 
 The code used in this work, found at \url{ltfat.org/notes/057}, illustrates the construction of several wavelet filter banks, as well as the use of some of the functions outlined above. A more detailed demo will be  integrated into the next LTFAT release. } 

\section{Experiments in Audio Processing}
\label{sec:experiments} 

We present three applications of the proposed wavelet decimation scheme. In the first application, we replicate experiments from prior work on NMF-based signal decomposition~\cite{fevotte2009nonnegative}, which originally relied on the STFT.
We simply substitute the wavelet transform for the STFT in a plug-and-play manner, leaving all other parameters unchanged. Only the parameters of the wavelet transform, i.e., the mother wavelet $\psi$, the number of channels $M$ and \revised{compensation} channels $M_{\textrm{C}}$, were adjusted. 
Given that the considered processing scheme was originally conceived for the STFT, and not adapted in any way, it is not expected that using the wavelet transform will outperform the original methods using the STFT.
Instead, the goal is to demonstrate that even a naive plug-and-play approach can achieve comparable results. 

\revised{In the second application, we consider onset detection in musical signals. 
Using the idea of spectral flux \cite[Sec.~3-A]{bello2005tutorial} for the wavelet coefficients in the proposed decimation scheme instead of the commonly used STFT coefficients, we calculate a detection function for onsets. 
The results for a small annotated test set \cite{onset_data} are compared to onsets detected by the same method using STFT coefficients as well as the output of a readily available onset detector \cite{mirtoolbox}.  
}

In a third application, we compare the performance of phaseless reconstruction with the fast Griffin-Lim algorithm (FGLA)~\cite{ltfatnote021,griffin1984signal} between the proposed decimation scheme and classical decimation. \revised{Since FGLA does not rely on the segmentation of the transform coefficients into time frames, this allows us to compare the processing  performance of our method to classical wavelet decimation. Additionally, results on phaseless reconstruction with FGLA from STFT coefficients are provided as a reference. Note that real-time variants of the Griffin-Lim algorithm do rely on frame-wise processing, e.g.,~\cite{Zhu2007realtime}, and could be adapted to the proposed decimation scheme.}

Accompanying audio files and code for reproducing the presented experiments are available at: \url{ltfat.org/notes/057}.

\subsection{Signal Decomposition with Nonnegative Matrix Factorization}

In \cite{fevotte2009nonnegative}, F\'evotte et al.\ showed that nonnegative matrix factorization with the Itakura-Saito cost function achieves a meaningful decomposition of STFT spectrograms of audio data as $S = W\cdot H$, with $W$ and $H$ being nonnegative component and activation matrices, respectively. 
The authors presented the decomposition, denoising, and upmixing of a 1920s recording of ``My Heart (Will Always Lead Me Back to You)'' by Louis Armstrong and His Hot Five. According to the subjective analysis in \cite{fevotte2009nonnegative}, the recording contains trumpet, clarinet, trombone, double bass, and piano tracks, as well as significant hiss noise and crackling. The signal used was an excerpt of length $108$\,s, sampled at $\xi_\textrm{samp} = 11\,025$\,Hz, with a total length of $L=1\,191\,735$ samples. \revised{In the original contribution, an Itakura-Saito NMF decomposition with $10$ components was computed from an STFT spectrogram at oversampling rate $2$, with $M=129$ channels from the zero to Nyquist frequencies. An inverse Gamma prior was used to regularize the component activation matrix $H$.}

\begin{figure*}[!t]
\vspace{-2pt}
\begin{center}
\includegraphics[width=0.9\textwidth,trim=0cm 1.7cm 0cm 1.3cm, clip]{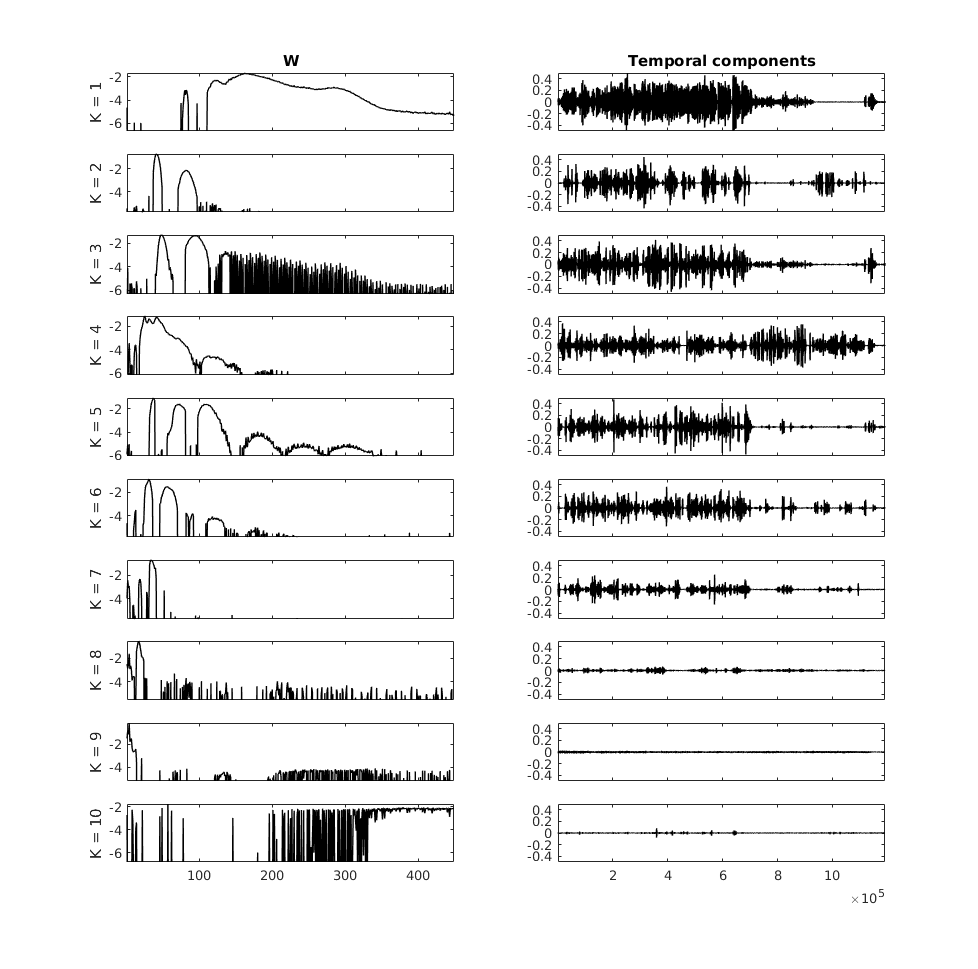}
\end{center}
\vspace{-8pt}
\caption{NMF decomposition of music excerpt. (Left) dB-scaled columns of $W$, i.e., NMF components. (Right) Reconstructed signal components. Components $4$, and $6$--$8$ capture most of the piano and double bass tracks. Components $9$, $10$ characterize large parts of the hiss and crackling noise. The remaining components contain most of the trumpet and clarinet lead track, as well as the trombone. \label{fig:nmf_results}}
\end{figure*}

In our experiment, we adapt the code provided with the follow-up contribution by F\'evotte~\cite{5946898}, which uses a different smoothing prior, but is otherwise identical. As in that contribution, the regularization parameter is set to $\lambda=25$. 
We use a wavelet transform with a Cauchy wavelet~\cite{ltfatnote053,dapa88}, with hyperparameter $\alpha=450$. Decimation is based on the Kronecker sequence and the oversampling rate is set to $2$. The parameters are optimized\footnote{Due to memory constraints, we restrict the optimization to choose no more than $769$ channels.} as in Section \ref{ssec:numeval}, leading to a total of $M+1=449$ channels and $M_{\textrm{C}}=6$ \revised{compensation} channels, i.e., the center frequency of $\psi_{l,0}$ is at $74$\,Hz. To achieve the desired oversampling rate, we set the decimation factor to $d=448$. The resulting NMF components and signal decomposition are shown in Fig.~\ref{fig:nmf_results}.

\vspace{6pt}
\noindent\textit{Results: }
Comparing to the results shown in \cite{fevotte2009nonnegative}, we see that the harmonic structure of the components in $W$ is not as pronounced when the wavelet transform is used, but still present. This is not entirely surprising, considering the large bandwidth of high frequency wavelets and the non-aligned phase space covering of the Kronecker sequence based decimation. Nonetheless, the decomposition achieves a separation into lead, accompaniment, and noise comparable to the results presented for the STFT in \cite{fevotte2009nonnegative,5946898}. The main difference is that the trombone track is not as clearly separated, but mostly mixed with the lead track. Audio examples, including individual components, as well as denoised and upmixed versions of the original signal, are provided on the website.

\begin{figure}[!t]
 \centering
  \includegraphics[width=0.44\textwidth,trim=0cm 0cm 0cm 0cm, clip]{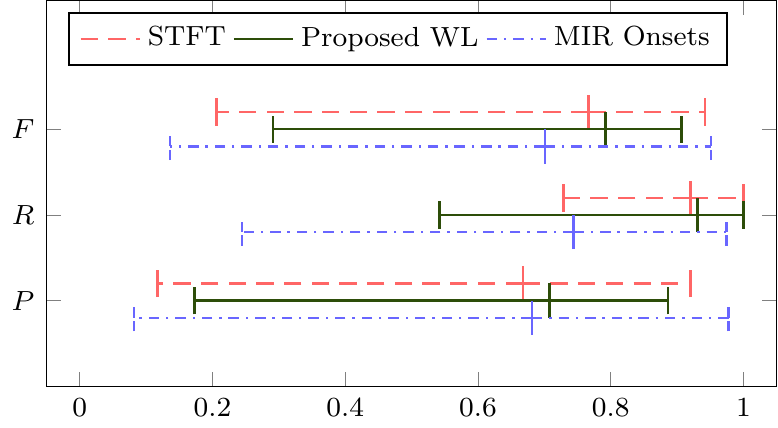}
\caption{
Whisker plots showing the minimal, median, and maximal F-measure $F$, recall $R$, and precision $P$, respectively, for 19 signals and three onset detectors.
The first two onset detectors are based on spectral flux calculated for the proposed wavelet transform or an STFT, respectively. The third method is the standard method from the MIRtoolbox based on an amplitude envelope. \label{fig:onsets}}
\end{figure}

\revised{
\subsection{Onset Detection in Music}
In our second experiment, we use the proposed wavelet decimation for the detection of onsets in audio signals.
More specifically, we use a spectral flux method as, e.g., described in \cite[Sec.~3-A]{bello2005tutorial} but replace the STFT with the proposed wavelet system. 
Spectral flux measures the increase in magnitude or energy in different frequency bands  (a decrease is set to 0) and takes the sum of these increments at each time frame.
The resulting time dependent spectral flux function is then used as a basis for a peak-picking procedure based on the assumption that local maxima of spectral flux are onsets of new musical events.

In our experiment, we use in place  of the  frequency channels of an STFT the channels of our proposed wavelet system. 
The resulting spectral flux for a signal $f$ is defined as
\begin{equation}
\label{eq:specflux}
    S(l) = \sum_j H\big(\abs{W_{\psi}f(x_{l,j},s_j)} - \abs{W_{\psi}f(x_{l-1,j},s_j)}\big)
\end{equation}
where $H(x)=(x+\abs{x})/2$ is a rectifier and the sum is only over the channels corresponding to our novel decimation scheme ignoring the $M_\textrm{C}$ compensation channels.
We choose a system with oversampling factor $4$, a Cauchy wavelet with $\alpha=2700$, and the number of channels resulting in the best frame bound, i.e., $M=1012$ and $M_{\textrm{C}}=20$ (see~Table~\ref{tab:framebounds}).
To avoid spurious local maxima, an additional time-dependent thresholding step is implemented.
We use here a multiple $\lambda$ of the local median which has been suggested as a robust choice in \cite[Sec.~4-B]{bello2005tutorial}.

For our experiment, we use the annotated onset detection database provided by the Pattern Recognition and Artificial Intelligence Group - University of Alicante (PRAIg-UA) 
\cite{onset_data}.
In the performance evaluation, we consider an onset to be correctly detected if the estimated onset is within 50ms of the annotated onset (this is a common measure in the literature \cite[Sec. 5-A]{bello2005tutorial}).
We then calculate the precision $P$ as the quotient of correctly detected onsets and total estimated onsets, and the recall $R$ as the quotient of correctly detected onsets and total annotated onsets. 
As a single performance measure, we further calculate the F-measure as 
$F=2 P R/(P+R)$.
We compare our method with spectral flux based on the STFT with a Hann window, i.e., replacing the wavelet coefficients in \eqref{eq:specflux} with STFT coefficients.
Here, we use the same decimation factor as in the wavelet case and adapt the number of channels and window length to obtain a tight frame with oversampling factor $4$.
Based on an optimization of the F-measure for the first audio sample in \cite{onset_data}, we choose the threshold factor $\lambda$ to be $1.24$ for the STFT and $1.34$ for the wavelet case.
This sample is excluded in the results below.
We compare our results to the basic onset detection algorithm  implemented in the MIRtoolbox \cite{mirtoolbox} which is based on an amplitude envelope.

\vspace{6pt}
\noindent\textit{Results: }
Our results are illustrated in Fig.~\ref{fig:onsets}.
The methods based on spectral flux are on average slightly better than the reference method based on an amplitude envelope. However, no method turned out to be universally best or worst over all signals.  
The difference between the STFT-based and the wavelet-based spectral flux is on average very small although for specific signals quite substantial differences are observed. 
This hints at the possibility that based on the signal class either method might be superior and a detailed analysis of their respective benefits and drawbacks is an interesting direction for further research. 
Finally, in both spectral flux cases the comparatively large recall and small precision suggests that the threshold factor $\lambda$ was actually chosen too small for a good balance between wrong and missed detections. }

\subsection{Phaseless Reconstruction with the Fast Griffin-Lim Algorithm}

In this experiment, we evaluate phaseless reconstruction, i.e., the reconstruction of an audio signal from magnitude-only time-frequency coefficients in our proposed wavelet decimation scheme. The Griffin-Lim algorithm~\cite{griffin1984signal} (GLA) remains the most popular iterative method for phaseless reconstruction from STFT or general time-frequency spectrograms. Here, we consider the fast Griffin-Lim (FGLA) variant proposed by Perraudin et al.~\cite{ltfatnote021}, which introduces a Nesterov-like acceleration term. The reconstruction error is measured as relative spectral error, often referred to as spectral convergence and given by 
\begin{equation}
\mathbf{err_{MS}}(f, f_r) = 10 ~ \log_{10} \frac{\Vert  \abs{W_\psi f_r} - \abs{W_\psi f} \Vert ^2}{\Vert W_\psi f \Vert ^2},
\label{eq:SNRformula}
\end{equation}
with the target signal $f$ and the reconstructed signal $f_r$. In the literature, it is not always clear with respect to which time-frequency representation the quantity $\mathbf{err_{MS}}$ is computed. Since we compare results across different representations and parameter choices, we fix the representation for computing $\mathbf{err_{MS}}$, as in \cite{Marafioti2021}. Here, we choose a highly oversampled wavelet transform using a Cauchy-type mother wavelet with $\alpha=1000$, with $M+1=181$ geometrically spaced frequency channels and $d=7$, i.e., approximately $50$-fold oversampling, as a reference representation. 

The experimental setup is similar to \cite[Section 4.2]{ltfatnote055}: We consider the same 15 signals from the EBU SQAM dataset\footnote{The first 5 seconds of signals 01, 02, 04, 14, 15, 16, 27, 39, 49, 50, 51, 52, 53, 54, and 70.} and test Cauchy-type mother wavelets with $\alpha=1000$, at oversampling rates\footnote{Note that the value $M/a$ considered in \cite[Section 4.2]{ltfatnote055} corresponds to roughly half the oversampling rate, considering that coefficients are complex-valued.} of $3$, $5$, and $10$. For the proposed method, we choose $M=750$, $968$, and $1369$ for low, medium, and high oversampling, respectively.
We further use $11$, $13$, and $17$ \revised{compensation} channels.
The number of \revised{compensation} channels is determined as in Section~\ref{ssec:numeval}. After promising results were obtained for $M=750$ at oversampling rate $3$, the parameters for higher rates are chosen by isotropic scaling of the sampling grid, i.e., the product $dM$ remains constant.%
\footnote{Note that the requirement that $d,M\in\NN$ leads to round-off errors. Moreover, the implementation we use chooses the largest value $d$ such that $M/(2d)$ is at least as large as the desired oversampling rate. Hence, the actual oversampling rate can be slightly larger.}
As references, we first consider wavelet transforms at the same oversampling rates, but with geometric frequency spacing. Since the uniform decimation scheme used in \cite[Section 4.2]{ltfatnote055} leads to unstable systems at low and medium oversampling, we further use channel-dependent decimation as discussed in Section II-A. The number of channels is set to $M=90$, $125$, and $180$ for low, medium, and high oversampling, respectively. 
Furthermore, we consider STFTs\revised{, with a $1536$ sample Hann window, }at the same redundancies with $M=1536$, $1920$, and $2880$ channels and uniform decimation.
To prevent issues with poor initialization, we first compute $20$ FGLA iterations for initial phase $0$ and for five random uniformly distributed phase initializations. 
The best of these six candidates is used to compute the final solution by applying another $130$ FGLA iterations. 

\begin{figure}[!t]
\begin{center}
\includegraphics[width=0.48\textwidth,trim=0cm 0cm 0cm 0cm, clip]{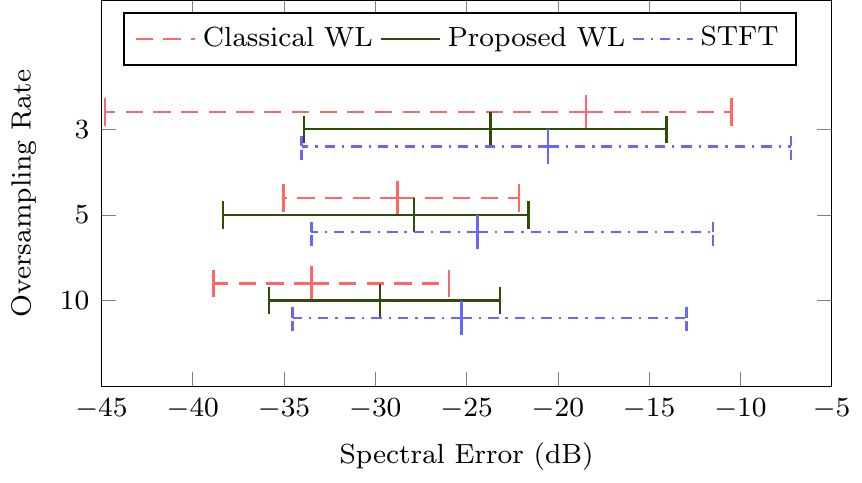}
\end{center}
\vspace{-4pt}
\caption{Whisker plots showing the minimal, median, and maximal spectral \revised{error} after phaseless reconstruction for 15 signals and three transforms, namely, the classical wavelet transform, the proposed wavelet transform with Kronecker sequence based decimation, and the STFT. The different oversampling rates are arranged vertically. \label{fig:pr_results}}
\end{figure}

\vspace{6pt}
\noindent\textit{Results: }
In Fig.~\ref{fig:pr_results}, our results are summarized. 
Overall, we see a clear trend that an increase in oversampling improves performance, matching the results obtained in \cite{Marafioti2021}. 
On average, the classical wavelet decimation is superior to the STFT at medium and high oversampling and to the proposed method at high oversampling rates. At medium oversampling both wavelet methods are roughly on par.  
At low oversampling, the performance of the proposed method is superior to both reference methods. On the associated website, we provide audio examples and additional results obtained when computing $\mathbf{err_{MS}}$ with respect to a reference STFT. As observed in~\cite{Marafioti2021}, $\mathbf{err_{MS}}$ is biased towards representations that are similar to the reference, and we found that $\mathbf{err_{MS}}$ with STFT reference, when compared to Fig.~\ref{fig:pr_results}, favors the STFT over the wavelet transform in general, and Kronecker decimation over classical wavelet decimation in particular. However, the overall findings are similar, such that we do not include these results here.

\section{Conclusion and Outlook}
\label{sec:conclude}

We proposed a novel, uniform wavelet decimation scheme with quasi-random delays. Conceptually, and through numerical evaluation, we demonstrated that the proposed scheme is suitable for constructing wavelet systems with the perfect reconstruction property, even at oversampling rates close to 1. In an audio decomposition application using Itakura-Saito NMF, we have shown that processing schemes previously proposed for the STFT are easily adapted to our setting by interpreting the filter bank coefficients as a time-frequency matrix, in contrast to classical, non-uniform wavelet decimation.  \revised{
Furthermore, we illustrated on a small test set of audio samples that our method can be used for onset detection by a straightforward adaptation of the spectral flux method originally defined for the STFT.}
Finally, we observed promising performance of our method in phaseless reconstruction for diverse audio signals.

We expect that the proposed decimation strategy will become a valuable asset for future work in different applications in audio and beyond. Furthermore, whereas the analysis of the proposed method in this work is restricted \revised{to finite systems, }\revthree{we expect the proposed construction to yield invertible wavelet transforms in the continuous domain as well. A rigorous study of its formal, mathematical properties in that setting is in preparation, considering frame and function space theory.}

\section*{Acknowledgments} 
We would like to thank C\'edric F\'evotte for kindly providing their code for Itakura-Saito NMF and some guidance on its use. We further thank Georg Taub\"ock for fruitful discussion on potential applications of the proposed method.

\begin{thebibliography}{10}
\providecommand{\url}[1]{#1}
\csname url@samestyle\endcsname
\providecommand{\newblock}{\relax}
\providecommand{\bibinfo}[2]{#2}
\providecommand{\BIBentrySTDinterwordspacing}{\spaceskip=0pt\relax}
\providecommand{\BIBentryALTinterwordstretchfactor}{4}
\providecommand{\BIBentryALTinterwordspacing}{\spaceskip=\fontdimen2\font plus
\BIBentryALTinterwordstretchfactor\fontdimen3\font minus
  \fontdimen4\font\relax}
\providecommand{\BIBforeignlanguage}[2]{{%
\expandafter\ifx\csname l@#1\endcsname\relax
\typeout{** WARNING: IEEEtran.bst: No hyphenation pattern has been}%
\typeout{** loaded for the language `#1'. Using the pattern for}%
\typeout{** the default language instead.}%
\else
\language=\csname l@#1\endcsname
\fi
#2}}
\providecommand{\BIBdecl}{\relax}
\BIBdecl

\bibitem{brown1992efficient}
J.~C. Brown and M.~S. Puckette, ``An efficient algorithm for the calculation of
  a constant {Q} transform,'' \emph{The Journal of the Acoustical Society of
  America}, vol.~92, no.~5, pp. 2698--2701, 1992.

\bibitem{moore2012introduction}
B.~C. Moore, \emph{An Introduction to the Psychology of Hearing}.\hskip 1em
  plus 0.5em minus 0.4em\relax Brill, 2012.

\bibitem{schorkhuber2010constant}
C.~Sch{\"o}rkhuber and A.~Klapuri, ``Constant-{Q} transform toolbox for music
  processing,'' in \emph{7th Sound and Music Computing Conference, Barcelona,
  Spain}, 2010, pp. 3--64.

\bibitem{schoerkhuber2013audio}
\BIBentryALTinterwordspacing
C.~Schörkhuber, A.~Klapuri, and A.~Sontacchi, ``Audio pitch shifting using the
  constant-{Q} transform,'' \emph{J. Audio Eng. Soc}, vol.~61, no. 7/8, pp.
  562--572, 2013. [Online]. Available:
  \url{http://www.aes.org/e-lib/browse.cfm?elib=16871}
\BIBentrySTDinterwordspacing

\bibitem{fuentesCQT2012}
B.~Fuentes, A.~Liutkus, R.~Badeau, and G.~Richard, ``Probabilistic model for
  main melody extraction using constant-{Q} transform,'' in \emph{2012 IEEE
  International Conference on Acoustics, Speech and Signal Processing
  (ICASSP)}, 2012, pp. 5357--5360.

\bibitem{TodiscoCQT2017}
\BIBentryALTinterwordspacing
M.~Todisco, H.~Delgado, and N.~Evans, ``Constant {Q} cepstral coefficients: A
  spoofing countermeasure for automatic speaker verification,'' \emph{Computer
  Speech \& Language}, vol.~45, pp. 516--535, 2017. [Online]. Available:
  \url{https://www.sciencedirect.com/science/article/pii/S0885230816303114}
\BIBentrySTDinterwordspacing

\bibitem{SAKAR2019255}
\BIBentryALTinterwordspacing
C.~O. Sakar, G.~Serbes, A.~Gunduz, H.~C. Tunc, H.~Nizam, B.~E. Sakar,
  M.~Tutuncu, T.~Aydin, M.~E. Isenkul, and H.~Apaydin, ``A comparative analysis
  of speech signal processing algorithms for {P}arkinson’s disease
  classification and the use of the tunable {Q}-factor wavelet transform,''
  \emph{Applied Soft Computing}, vol.~74, pp. 255--263, 2019. [Online].
  Available:
  \url{https://www.sciencedirect.com/science/article/pii/S1568494618305799}
\BIBentrySTDinterwordspacing

\bibitem{119752}
S.~Kadambe and G.~Boudreaux-Bartels, ``Application of the wavelet transform for
  pitch detection of speech signals,'' \emph{IEEE Transactions on Information
  Theory}, vol.~38, no.~2, pp. 917--924, 1992.

\bibitem{1021072}
G.~Tzanetakis and P.~Cook, ``Musical genre classification of audio signals,''
  \emph{IEEE Transactions on Speech and Audio Processing}, vol.~10, no.~5, pp.
  293--302, 2002.

\bibitem{1495445}
C.-C. Lin, S.-H. Chen, T.-K. Truong, and Y.~Chang, ``Audio classification and
  categorization based on wavelets and support vector machine,'' \emph{IEEE
  Transactions on Speech and Audio Processing}, vol.~13, no.~5, pp. 644--651,
  2005.

\bibitem{Mallat2009Wavelet}
S.~Mallat, \emph{A Wavelet Tour of Signal Processing (Third Edition)}.\hskip
  1em plus 0.5em minus 0.4em\relax Boston: Academic Press, 2009.

\bibitem{10.1007/978-3-642-75988-8_28}
M.~Holschneider, R.~Kronland-Martinet, J.~Morlet, and P.~Tchamitchian, ``A
  real-time algorithm for signal analysis with the help of the wavelet
  transform,'' in \emph{Wavelets}, J.-M. Combes, A.~Grossmann, and
  P.~Tchamitchian, Eds.\hskip 1em plus 0.5em minus 0.4em\relax Berlin,
  Heidelberg: Springer Berlin Heidelberg, 1990, pp. 286--297.

\bibitem{10.1007/978-3-642-75988-8_29}
P.~Dutilleux, ``An implementation of the ``algorithme {\`a} trous'' to compute
  the wavelet transform,'' in \emph{Wavelets}, J.-M. Combes, A.~Grossmann, and
  P.~Tchamitchian, Eds.\hskip 1em plus 0.5em minus 0.4em\relax Berlin,
  Heidelberg: Springer Berlin Heidelberg, 1990, pp. 298--304.

\bibitem{SelesnickWavelet2011}
I.~W. Selesnick, ``Wavelet transform with tunable {Q}-factor,'' \emph{IEEE
  Transactions on Signal Processing}, vol.~59, no.~8, pp. 3560--3575, 2011.

\bibitem{dogrhove13}
N.~Holighaus, M.~D{\"o}rfler, G.~A. Velasco, and T.~Grill, ``{A} framework for
  invertible, real-time constant-{Q} transforms,'' \emph{IEEE Audio, Speech,
  Language Process.}, vol.~21, no.~4, pp. 775--785, Apr. 2013.

\bibitem{schoerkhuber2014a}
\BIBentryALTinterwordspacing
C.~Schörkhuber, A.~Klapuri, N.~Holighaus, and M.~Dörfler, ``A matlab toolbox
  for efficient perfect reconstruction time-frequency transforms with
  log-frequency resolution,'' in \emph{Proceedings of the 53rd International
  Audio Engineering Society Conference: Semantic Audio}, Jan 2014. [Online].
  Available: \url{http://www.aes.org/e-lib/browse.cfm?elib=17112}
\BIBentrySTDinterwordspacing

\bibitem{Christensen2016}
O.~Christensen, \emph{An Introduction to Frames and Riesz Bases}, ser. Applied
  and Numerical Harmonic Analysis.\hskip 1em plus 0.5em minus 0.4em\relax Cham:
  Springer International Publishing, 2016.

\bibitem{pd}
M.~S. Puckette \emph{et~al.}, ``Pure data,'' in \emph{ICMC}, 1997.

\bibitem{mirtoolbox}
O.~Lartillot, P.~Toiviainen, and T.~Eerola, ``{A {M}atlab Toolbox for Music
  Information Retrieval},'' in \emph{Data Analysis, Machine Learning and
  Applications}, C.~Preisach, H.~Burkhardt, L.~Schmidt-Thieme, and R.~Decker,
  Eds.\hskip 1em plus 0.5em minus 0.4em\relax Berlin, Germany: Springer, 2008,
  pp. 261--268.

\bibitem{mp3standard}
``Information technology — coding of moving pictures and associated audio for
  digital storage media at up to about 1,5 mbit/s — part 3: Audio,''
  International Organization for Standardization, Geneva, CH, Standard, Aug.
  1993.

\bibitem{gardner}
W.~G. Gardner, ``Efficient convolution without input/output delay,'' in
  \emph{Audio Engineering Society Convention 97}.\hskip 1em plus 0.5em minus
  0.4em\relax Audio Engineering Society, 1994.

\bibitem{necciari18}
T.~Necciari, N.~Holighaus, P.~Balazs, Z.~Pr\r{u}\v{s}a, P.~Majdak, and
  O.~Derrien, ``Audlet filter banks: A versatile analysis/synthesis framework
  using auditory frequency scales,'' \emph{Applied Sciences}, vol.~8, no. 1:96,
  2018.

\bibitem{6637697}
T.~Necciari, P.~Balazs, N.~Holighaus, and P.~L. Søndergaard, ``The erblet
  transform: An auditory-based time-frequency representation with perfect
  reconstruction,'' in \emph{2013 IEEE International Conference on Acoustics,
  Speech and Signal Processing}, 2013, pp. 498--502.

\bibitem{balazs2011theory}
P.~Balazs, M.~D{\"o}rfler, F.~Jaillet, N.~Holighaus, and G.~Velasco, ``Theory,
  implementation and applications of nonstationary {G}abor frames,'' \emph{J.
  Comput. Appl. Math.}, vol. 236, no.~6, pp. 1481--1496, Oct. 2011.

\bibitem{levie2021randomized1}
R.~Levie and H.~Avron, ``Randomized signal processing with continuous frames,''
  \emph{J Fourier Anal Appl}, vol.~28, no.~5, 2021.

\bibitem{levie2021randomized}
------, ``Randomized continuous frames in time-frequency analysis,''
  \emph{arXiv preprint arXiv:2009.10525 [math.NA]}, 2021.

\bibitem{levie2021quasi}
R.~Levie, H.~Avron, and G.~Kutyniok, ``Quasi {M}onte {C}arlo time-frequency
  analysis,'' \emph{arXiv preprint arXiv:2011.02025 [math.NA]}, 2021.

\bibitem{bass2013relevant}
\BIBentryALTinterwordspacing
R.~F. Bass and K.~Gröchenig, ``{Relevant sampling of band-limited
  functions},'' \emph{Illinois Journal of Mathematics}, vol.~57, no.~1, pp. 43
  -- 58, 2013. [Online]. Available:
  \url{https://doi.org/10.1215/ijm/1403534485}
\BIBentrySTDinterwordspacing

\bibitem{FUHR20191}
H.~Führ and J.~Xian, ``Relevant sampling in finitely generated shift-invariant
  spaces,'' \emph{Journal of Approximation Theory}, vol. 240, pp. 1--15, 2019.

\bibitem{PATEL2020124270}
D.~Patel and S.~Sampath, ``Random sampling in reproducing kernel subspaces of
  $\mathbf l^p(\mathbb{R}^n)$,'' \emph{Journal of Mathematical Analysis and
  Applications}, vol. 491, no.~1, p. 124270, 2020.

\bibitem{goyal2021random}
P.~Goyal, D.~Patel, and S.~Sampath, ``Random sampling in reproducing kernel
  subspace of mixed lebesgue spaces,'' 2021.

\bibitem{velasco2017relevant}
G.~A. Velasco, ``Relevant sampling of the short-time fourier transform of
  time-frequency localized functions,'' 2017.

\bibitem{Janssen1998}
A.~J. E.~M. Janssen, \emph{The duality condition for Weyl-Heisenberg
  frames}.\hskip 1em plus 0.5em minus 0.4em\relax Boston, MA: Birkh{\"a}user
  Boston, 1998, pp. 33--84.

\bibitem{BoelcskeiFilterbanks1998}
H.~B\"olcskei, F.~Hlawatsch, and H.~Feichtinger, ``Frame-theoretic analysis of
  oversampled filter banks,'' \emph{IEEE Transactions on Signal Processing},
  vol.~46, no.~12, pp. 3256--3268, 1998.

\bibitem{Niederreiter1992random}
\BIBentryALTinterwordspacing
H.~Niederreiter, \emph{Random Number Generation and Quasi-{M}onte {C}arlo
  Methods}.\hskip 1em plus 0.5em minus 0.4em\relax Society for Industrial and
  Applied Mathematics, 1992. [Online]. Available:
  \url{https://epubs.siam.org/doi/abs/10.1137/1.9781611970081}
\BIBentrySTDinterwordspacing

\bibitem{dick_pillichshammer_2010}
J.~Dick and F.~Pillichshammer, \emph{Digital Nets and Sequences: Discrepancy
  Theory and Quasi–{M}onte {C}arlo Integration}.\hskip 1em plus 0.5em minus
  0.4em\relax Cambridge University Press, 2010.

\bibitem{fevotte2009nonnegative}
C.~Févotte, N.~Bertin, and J.-L. Durrieu, ``Nonnegative matrix factorization
  with the {I}takura-{S}aito divergence: With application to music analysis,''
  \emph{Neural Computation}, vol.~21, no.~3, pp. 793--830, 2009.

\bibitem{bello2005tutorial}
J.~P. Bello, L.~Daudet, S.~Abdallah, C.~Duxbury, M.~Davies, and M.~B. Sandler,
  ``A tutorial on onset detection in music signals,'' \emph{IEEE Speech Audio
  Process.}, vol.~13, no.~5, pp. 1035--1047, 2005.

\bibitem{ltfatnote021}
N.~Perraudin, P.~Balazs, and P.~L. Sondergaard, ``A fast {G}riffin-{L}im
  algorithm,'' in \emph{Proc. IEEE Appl. Sig. Process. Audio Acoustics}, New
  Paltz, NY, USA, Oct. 2013.

\bibitem{ltfatnote053}
N.~Holighaus, G.~Koliander, Z.~Průša, and L.~D. Abreu, ``Characterization of
  analytic wavelet transforms and a new phaseless reconstruction algorithm,''
  \emph{IEEE Transactions on Signal Processing}, vol.~67, no.~15, pp.
  3894--3908, 2019.

\bibitem{Lilly2010analytic}
J.~M. Lilly and S.~C. Olhede, ``On the analytic wavelet transform,'' \emph{IEEE
  Transactions on Information Theory}, vol.~56, no.~8, pp. 4135--4156, 2010.

\bibitem{dagrme86}
I.~Daubechies, A.~Grossmann, and Y.~Meyer, ``Painless nonorthogonal
  expansions,'' \emph{Journal of Mathematical Physics}, vol.~27, no.~5, pp.
  1271--1283, 1986.

\bibitem{gr93acceleration}
K.~Grochenig, ``Acceleration of the frame algorithm,'' \emph{IEEE Transactions
  on Signal Processing}, vol.~41, no.~12, pp. 3331--3340, 1993.

\bibitem{da92}
I.~{D}aubechies, \emph{{T}en lectures on wavelets.}\hskip 1em plus 0.5em minus
  0.4em\relax {S}{I}{A}{M}, 1992.

\bibitem{DT97}
M.~Drmota and R.~F. Tichy, \emph{Sequences, Discrepancies and
  Applications}.\hskip 1em plus 0.5em minus 0.4em\relax Springer Berlin
  Heidelberg, 1997.

\bibitem{kuinie}
L.~Kuipers and H.~Niederreiter, \emph{Uniform Distribution of Sequences}.\hskip
  1em plus 0.5em minus 0.4em\relax John Wiley, 1974.

\bibitem{vdc}
J.~G. {van der Corput}, ``\BIBforeignlanguage{German}{{Verteilungsfunktionen
  I-II}},'' \emph{\BIBforeignlanguage{German}{{Proc. Akad. Wet. Amsterdam}}},
  vol.~38, pp. 813--821, 1058–1066, 1935.

\bibitem{FAURE2015760}
H.~Faure, P.~Kritzer, and F.~Pillichshammer, ``From van der {C}orput to modern
  constructions of sequences for quasi-{M}onte {C}arlo rules,''
  \emph{Indagationes Mathematicae}, vol.~26, no.~5, pp. 760--822, 2015, in
  memoriam J.G. van der Corput (1890–1975).

\bibitem{dapa88}
I.~Daubechies and T.~Paul, ``Time-frequency localisation operators---a
  geometric phase space approach: {II} {T}he use of dilations,'' \emph{Inverse
  Prob.}, vol.~4, no.~3, pp. 661--680, Aug. 1988.

\bibitem{ltfatnote055}
\BIBentryALTinterwordspacing
N.~Holighaus, G.~Koliander, Z.~Pr\r{u}\v{s}a, and L.~Abreu, ``Non-iterative
  phaseless reconstruction from wavelet transform magnitude,'' in
  \emph{Proceedings of the International Conference on Digital Audio Effects
  2019 (DAFx19)}, Sept 2019. [Online]. Available:
  \url{http://dafx.de/paper-archive/2019/DAFx2019_paper_23.pdf}
\BIBentrySTDinterwordspacing

\bibitem{abreuAccuSpecs}
L.~D. Abreu, K.~Gröchenig, and J.~L. Romero, ``On accumulated spectrograms,''
  \emph{Transactions of the American Mathematical Society}, vol. 368, pp.
  3629--3649, 01 2016.

\bibitem{smith2009audio}
J.~O. Smith, ``Audio {FFT} filter banks,'' \emph{Proceedings of 12th
  International Conference on Digital Audio Effects (DAFx-09), Como}, 2009.

\bibitem{velasco2011constructing}
G.~A. Velasco, N.~Holighaus, M.~D{\"o}rfler, and T.~Grill, ``Constructing an
  invertible constant-{Q} transform with non-stationary {G}abor frames,''
  \emph{Proceedings of DAFX11, Paris}, vol.~33, 2011.

\bibitem{ltfatnote026}
Z.~Pr\r{u}\v{s}a, ``Segmentwise discrete wavelet transform,'' Ph.D.
  dissertation, Brno University of Technology, Brno, 2012.

\bibitem{daubechies88}
I.~Daubechies and T.~Paul, ``{Time-frequency localisation operators—a
  geometric phase space approach: II The use of dilations},'' \emph{{Inverse
  Problems}}, vol.~4, no.~3, pp. 661–--680, 1988.

\bibitem{olwa02}
S.~C. Olhede and A.~T. Walden, ``Generalized {M}orse wavelets,'' \emph{{IEEE}
  Trans. Sig. Process.}, vol.~50, no.~11, pp. 2661--2670, Nov. 2002.

\bibitem{gao2010wavelets}
R.~X. Gao and R.~Yan, \emph{Wavelets: Theory and applications for
  manufacturing}.\hskip 1em plus 0.5em minus 0.4em\relax Springer Science \&
  Business Media, 2010.

\bibitem{chaudhuryunser09}
K.~N. Chaudhury and M.~Unser, ``Construction of {H}ilbert {T}ransform {P}airs
  of {W}avelet {B}ases and {G}abor-{L}ike {T}ransforms,'' \emph{{IEEE} Trans.
  Sig. Process.}, vol.~57, no.~9, pp. 3411–--3425, 2009.

\bibitem{ltfatnote051}
Z.~Pr\r{u}\v{s}a and N.~Holighaus, ``Non-iterative filter bank phase
  (re)construction,'' in \emph{Proc. 25th European Signal Processing Conference
  (EUSIPCO--2017)}, Aug 2017, pp. 952--956.

\bibitem{ltfatnote041}
N.~Holighaus, Z.~Pr\r{u}\v{s}a, and P.~L.~S. ndergaard, ``Reassignment and
  synchrosqueezing for general time-frequency filter banks, subsampling and
  processing,'' \emph{{Signal Processing}}, vol. 125, no. Supplement C, pp. 1
  -- 8, 2016.

\bibitem{onset_data}
\BIBentryALTinterwordspacing
``Onset detection database,'' provided by the Pattern Recognition and
  Artificial Intelligence Group - University of Alicante (PRAIg-UA). [Online].
  Available: \url{https://grfia.dlsi.ua.es/cm/projects/prosemus/database.php}
\BIBentrySTDinterwordspacing

\bibitem{griffin1984signal}
D.~Griffin and J.~Lim, ``Signal estimation from modified short-time {F}ourier
  transform,'' \emph{IEEE Trans. Acoust., Speech, Signal Process.}, vol.~32,
  no.~2, pp. 236--243, Apr. 1984.

\bibitem{Zhu2007realtime}
X.~Zhu, G.~Beauregard, and L.~Wyse, ``Real-time signal estimation from modified
  short-time fourier transform magnitude spectra,'' \emph{IEEE Transactions on
  Audio Speech and Language Processing}, vol.~15, pp. 1645 -- 1653, 07 2007.

\bibitem{5946898}
C.~Févotte, ``Majorization-minimization algorithm for smooth {I}takura-{S}aito
  nonnegative matrix factorization,'' in \emph{2011 IEEE International
  Conference on Acoustics, Speech and Signal Processing (ICASSP)}, 2011, pp.
  1980--1983.

\bibitem{Marafioti2021}
A.~Marafioti, N.~Holighaus, and P.~Majdak, ``Time-frequency phase retrieval for
  audio—the effect of transform parameters,'' \emph{IEEE Transactions on
  Signal Processing}, vol.~69, pp. 3585--3596, 2021.

\end{thebibliography}


\end{document}